\documentclass{emulateapj}
\usepackage{apjfonts}
\usepackage{graphicx}
\usepackage{epsfig}
\usepackage{rotating}

\newcommand{\bc}{\begin{center}}
\newcommand{\ec}{\end{center}}
\newcommand{\be}{\begin{equation}}
\newcommand{\ee}{\end{equation}}
\newcommand{\ba}{\begin{eqnarray}}
\newcommand{\ea}{\end{eqnarray}}
\newcommand{\bt}{\begin{tabular}}
\newcommand{\et}{\end{tabular}}

\def\farcm{\hbox{$.\!\!^{\prime}$}}

\def\farcm{\hbox{$.\!\!^{\prime}$}}

\begin{document}
\submitted{Submitted to ApJ 2007 May 12}

\title{X-ray emission from PSR J1809--1917 and its pulsar wind nebula,
possibly associated with the TeV gamma-ray source HESS J1809--193}

\author{
 O.\ Kargaltsev and G.\ G.\ Pavlov}
\affil{The Pennsylvania State University, 525 Davey Lab, University
Park, PA 16802, USA} \email{oyk100@psu.edu,pavlov@astro.psu.edu}

\begin{abstract}

We detected X-ray emission from the 50-kyr-old pulsar J1809--1917
and resolved its pulsar wind nebula (PWN) with the {\sl Chandra X-ray
Observatory}. The pulsar's observed flux is $F_{\rm psr} = (1.8\pm
0.2)\times 10^{-14}$ ergs cm$^{-2}$ s$^{-1}$ in the 1--6 keV band.  A
two-component blackbody+power-law (BB+PL) fit of the pulsar's
spectrum yields the photon index $\Gamma_{\rm psr}=1.2\pm 0.6$
and luminosity $L_{\rm psr}=(4\pm 1)\times 10^{31}$ ergs s$^{-1}$ of
the PL component, in the 0.5--8 keV band, for a plausible distance
$d=3.5$ kpc and $n_{\rm H}=0.7\times 10^{22}$ cm$^{-2}$.
 The BB component corresponds to the temperature $T\approx 2$ MK,
and bolometric luminosity
 $L_{\rm bol}
\sim 1\times 10^{32}$ ergs s$^{-1}$. The bright inner PWN
component of a $3''\times 12''$ size is elongated in the north-south
 direction,
 with the
pulsar
close to its south
 end.  This component
is immersed in a larger ($\approx 20''\times 40''$), similarly elongated
outer PWN component of lower surface brightness. The elongated
shape of the compact PWN can be explained by the ram pressure
confinement of the pulsar wind due to the supersonic motion of the
pulsar. The observed flux of the compact PWN, including both
components, is $F_{\rm pwn}\simeq (1.5\pm 0.1)\times10^{-13}$
ergs cm$^{-2}$ s$^{-1}$ in the 1--6 keV band. The PWN spectrum can
be fitted with a PL model with $n_{\rm H}\approx 0.7 \times 10^{22}$
cm$^{-2}$ and photon index $\Gamma_{\rm pwn}=1.4\pm0.1$,
corresponding to the 0.5--8 keV luminosity $L_{\rm pwn} \approx
 4\times 10^{32}$
ergs s$^{-1}$.  The compact PWN appears to be inside a more
extended ($\approx4'\times4'$) emission with the total observed flux
$F_{\rm ext}\sim 5\times10^{-13}$ ergs s$^{-1}$ in the 0.8--7 keV
band. This large-scale emission is more extended to the south of the
pulsar, i.e.\ in the direction of the alleged pulsar motion. To explain
the extended X-ray emission
 ahead of the moving pulsar, one has to
invoke strong intrinsic anisotropy of the pulsar wind or assume that
this emission comes from a relic PWN swept by the asymmetrical
reverse SNR shock.
  The pulsar
and its PWN are located within the extent of the unidentified TeV
source HESS~J1809--193. The
brightest part of the TeV source is offset by
$\sim 8'$
 to the south of the pulsar,
i.e.\ in the same direction as the large-scale X-ray emission. Although
the
 association between the PSR~J1809--1917 and HESS~J1809--193
is plausible, an alternative source of relativistic electrons powering
HESS~J1809--193 might be the serendipitously discovered X-ray
source CXOU J180940.7$-$192544.
 In addition to the
CMBR or Galactic starlight background, the low-frequency seed
photons for Compton upscattering to TeV energies might be supplied
by bright infrared emission from dust-molecular clouds seen within
HESS~J1809--193.

\end{abstract}
\section{Introduction}

  {\sl Chandra} and {\sl XMM-Newton}  observations
  have
   established
 the ubiquity of
X-ray pulsar wind nebulae (PWNe) around young rotation-powered
pulsars (see the reviews by Kaspi et al.\ 2006 and Gaensler \& Slane
2006). The X-ray PWN emission is produced by relativistic particles
gyrating in the magnetic field downstream of the termination shock
 in the pulsar
wind (Kennel \& Coroniti 1994; Arons 2004).
Most of the PWNe have been discovered around young ($\tau\lesssim 30$  kyr)
pulsars.
 The innermost parts of the young PWNe often show
axisymmetric morphologies, including toroidal structures and jets
along the pulsar's spin axis. Recently, it has become apparent that
PWNe
 accompanying older pulsars
can also be quite luminous (e.g., McGowan et al.\ 2006). Many of
these older PWNe exhibit cometary morphologies
 indicating that the pulsar wind is confined by the ram pressure
caused by the supersonic motion of the pulsar in the ambient
medium. Studying X-ray PWNe of various ages helps understand the
nature and evolution of the ultrarelativistic pulsar winds and their
interaction with the ambient medium.

An interesting object for such investigations is PSR J809--1917
(hereafter J1809).  The discovery of this radio pulsar ($P=82.7$ ms) in
the Parkes Multibeam Pulsar Survey\footnote{
http://www.atnf.csiro.au/research/pulsar/pmsurv} was reported by
Morris et al.\ (2002). The pulsar's dispersion measure, ${\rm
DM}=197\, {\rm cm}^{-3}\, {\rm pc}$, and the Galactic electron
distribution models by Taylor \& Cordes  (1993) and Cordes \& Lazio
(2002) give the distance to the pulsar of 3.7 and 3.5 kpc, respectively.
Having the spin-down age $\tau\equiv P/2\dot{P} =51$ kyr and
spin-down power $\dot{E}\equiv 4\pi^2 I \dot{P} P^{-3} \simeq
1.8\times 10^{36} $ ergs s$^{-1}$, J1809 is somewhat older and less
energetic than the famous Vela pulsar ($\tau = 11$ kyr,
$\dot{E}=6.9\times 10^{36}$ ergs s$^{-1}$), which is accompanied by
a remarkable PWN resolved in radio (Dodson et al.\ 2003b) and
X-rays
   (Pavlov
et al.\ 2003, and references therein). However, it is much more
energetic than typical ``middle-aged'' pulsars, such as B0656+14
($\tau=110$ kyr, $\dot{E}=3.8\times 10^{34}$ ergs s$^{-1}$) and
Geminga ($\tau =340$ kyr, $\dot{E} =3.2\times 10^{34}$ ergs
s$^{-1}$),  whose PWNe are very faint (e.g., Pavlov et al.\ 2006).

J1809 is young enough to look for a remnant of the supernova that
created the pulsar. Deep radio observations by Brogan et al.\ (2004)
have
revealed two compact HII regions and two SNRs, G11.03$-$0.05 and
G11.18+0.11, projected near the pulsar (the pulsar's Galactic
coordinates are $l=11.094^\circ$, $b=+0.080^\circ$). The distances to
the SNRs, estimated from the radio surface brightness-diameter
($\Sigma-D$) relation (Case \& Bhattacharya 1998), are $\sim 16$ and
$\sim 17$ kpc, respectively. However, given the very large uncertainty
of the $\Sigma-D$ relation for faint SNRs,  the association of J1809
with one
 of the SNRs cannot be ruled out despite the discrepant distance estimates.
Brogan et al.\ (2004)
 estimated that the pulsar would need to have the transverse speed of
about 200 or
 140 km s$^{-1}$ to originate from the geometrical
  center of G11.03$-$0.05 or G11.18+0.11, respectively,
assuming that all three are at the same distance of
 4~kpc and the pulsar's true age is 50 kyrs.
The estimated speeds are close to the average speeds recently measured
for a large sample
 of radio pulsars by Hobbs et al.\ (2005).

 Before the detailed radio studies were carried out, the region
had been observed by {\sl ASCA}, first as  part of the
  Galactic plane survey (Sugizaki et al.\ 2001),
and then with a deeper follow-up exposure (Bamba et al.\ 2003;
Ueno et al.\ 2005).
   These observations revealed an amorphous diffuse emission that encompassed the pulsar position.
    No pulsar has been detected in these X-ray observations.
The  observed large-scale emission was attributed to a new SNR,
dubbed G11.0+0.0. Bamba et al.\  (2003) have
   found that the X-ray spectrum
 of the putative SNR fits
the absorbed power-law (PL) model with photon index
 $\Gamma=1.6^{+0.3}_{-0.2}$,
hydrogen column density $n_{\rm H,22}\equiv n_{\rm
H}/(10^{22}~{\rm cm}^{-2})=0.8\pm0.3$,
  and
0.7--10 keV flux
   of $3.8\times 10^{-12}$ ergs cm$^{-2}$ s$^{-1}$.
Using the best-fit $n_{\rm H}$ and assuming that the mean density in
the Galactic plane is 1 H cm$^{-3}$, Bamba et al.\  (2003)
     estimated the distance of  2.6 kpc,
 which gives the X-ray luminosity of $\approx3.7\times 10^{33}$ ergs s$^{-1}$ (in 0.7--10 keV).
  These authors suggested that G11.0+0.0 could be
a Crab-like (plerionic) SNR.

 Studying of the J1809 field has become particularly interesting after
the recent discovery of the TeV $\gamma$-ray source
HESS~J1809--193 (hereafter HESS\,J1809; Aharonian et al.\ 2007).  The
brightest, firmly detected part of this extended source (radius
$\sim12'$--$15'$) is centered at R.A.$ = 18^{\rm h}09.8^{\rm m}$,
decl.$=-19^{\circ}25'$, about $8'$ from J1809; its $\gamma$-ray flux
is $F_\gamma\approx 1.4\times 10^{-12}$ ergs cm$^{-2}$ s$^{-1}$ in
the 1--10 TeV band. Given the small angular separation, it seems
plausible that the HESS source could be powered by this pulsar, in
which case its $\gamma$-ray luminosity, $L_\gamma\approx 2\times
10^{34} d^{2}_{3.5}$ ergs s$^{-1}$, would be about 1\% of the pulsar's
spindown power.

In this paper, we describe the results of a {\sl Chandra} observation of
PSR J1809--1917 and its compact synchrotron nebula\footnote{
Preliminary results of this observation have been presented by
Sanwal et al.\ (2005).} and discuss its possible connection to
HESS~J1809. We also
 describe the multiwavelength  properties of objects located within
  the central part of HESS~J1809,
including two newly discovered X-ray sources, and discuss their
relation to HESS~J1809. The details of the observation and the data
analysis are presented in \S2. In \S3 we discuss  possible
interpretations of the PWN morphology, describe inferences from the
pulsar spectrum,  and speculate on the nature of HESS~J1809 and its
relation
 to the other sources in the field.
  Our main results are summarized
in \S4.

\section{Observations and Data Analysis}

 J1809 was observed with the Advanced CCD Imaging Spectrometer (ACIS)
on board {\sl Chandra} on 2004 July 21 (ObsID 3853).
 The useful scientific exposure time was 19,955 s. The
observation was carried out in Faint mode, and the pulsar was
imaged on S3 chip, $\approx0.72'$ off-axis. The other ACIS chips
activated during this observation were S1, S2, S4, I2, and I3. The
detector was operated in Full Frame mode, which provides time
resolution of 3.24 seconds. The data were reduced using the Chandra
Interactive Analysis of Observations (CIAO) software (ver.\ 3.2.1;
CALDB ver.\ 3.0.3).

\begin{figure*}[t]
 \centering
\includegraphics[width=6.5in,angle=0]{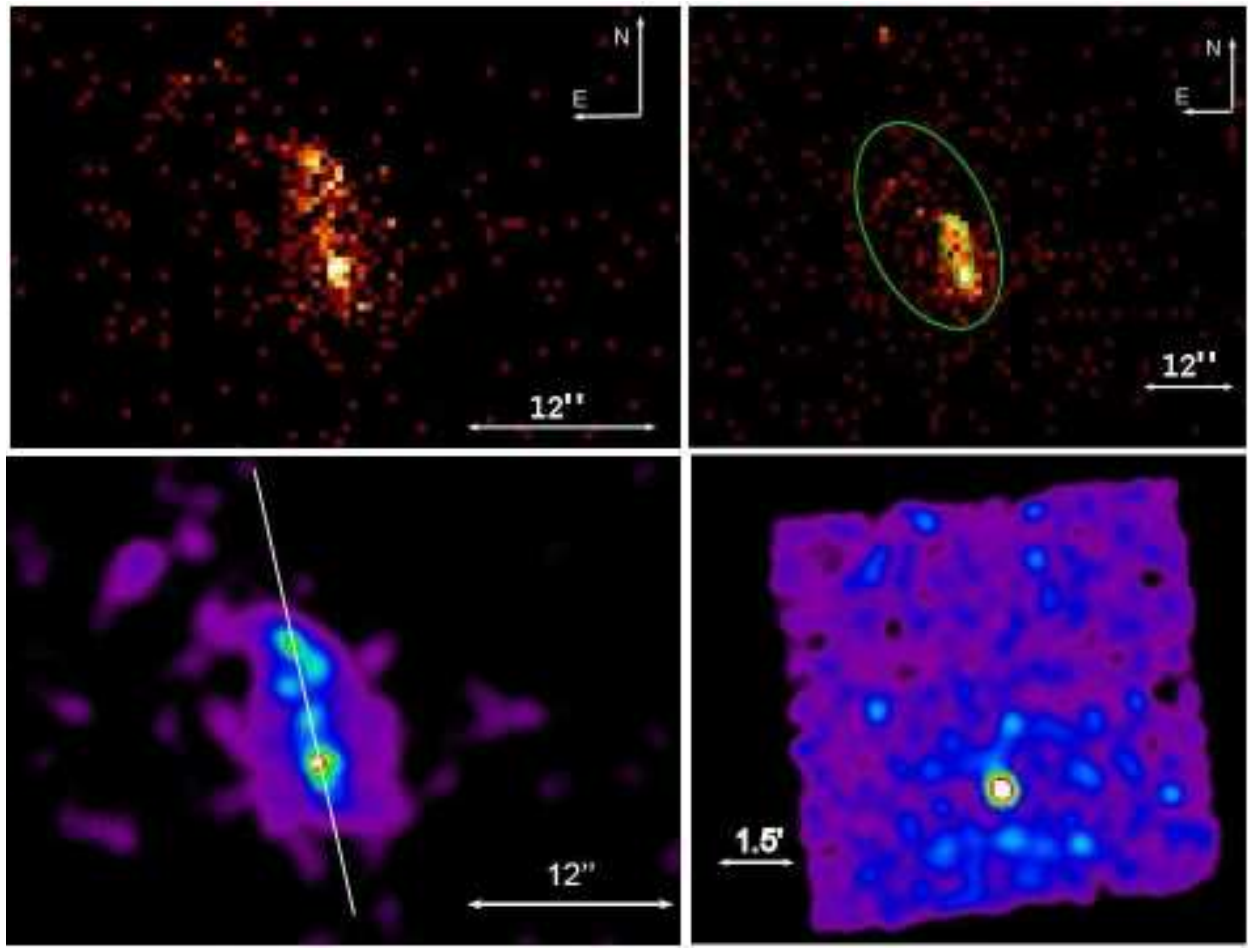}
 \caption{ {\em Top left:} ACIS-S3 image
 of J1809 and its PWN (0.8--7 keV; pixel size $0.49''$). {\em Top right:}
Extraction regions used for the analysis of the PWN components
(0.8--7 keV; pixel size $0.98''$).
 {\em Bottom left:} Adaptively smoothed sub-pixel resolution image
 (0.8--7 keV; pixel size $0.25''$)
obtained by removing the pipeline pixel-randomization and applying
the sub-pixel resolution tool (based on analyzing the charge
distribution produced by an X-ray event; Tsunemi et al.\ 2001; Mori et
al.\ 2001). The straight line shows the approximate symmetry axis of
the X-ray PWN. {\em Bottom right:}
 Heavily binned (pixels size $3.94''$) and
 smoothed (with a gaussian kernel of $r=25''$)
 ACIS-S3 image of J1809 and its PWN. The brightness and smoothing scales are chosen
 to show the fainter, more extended emission.
}
\end{figure*}

\begin{figure}
 \centering
 \includegraphics[width=2.5in,angle=0]{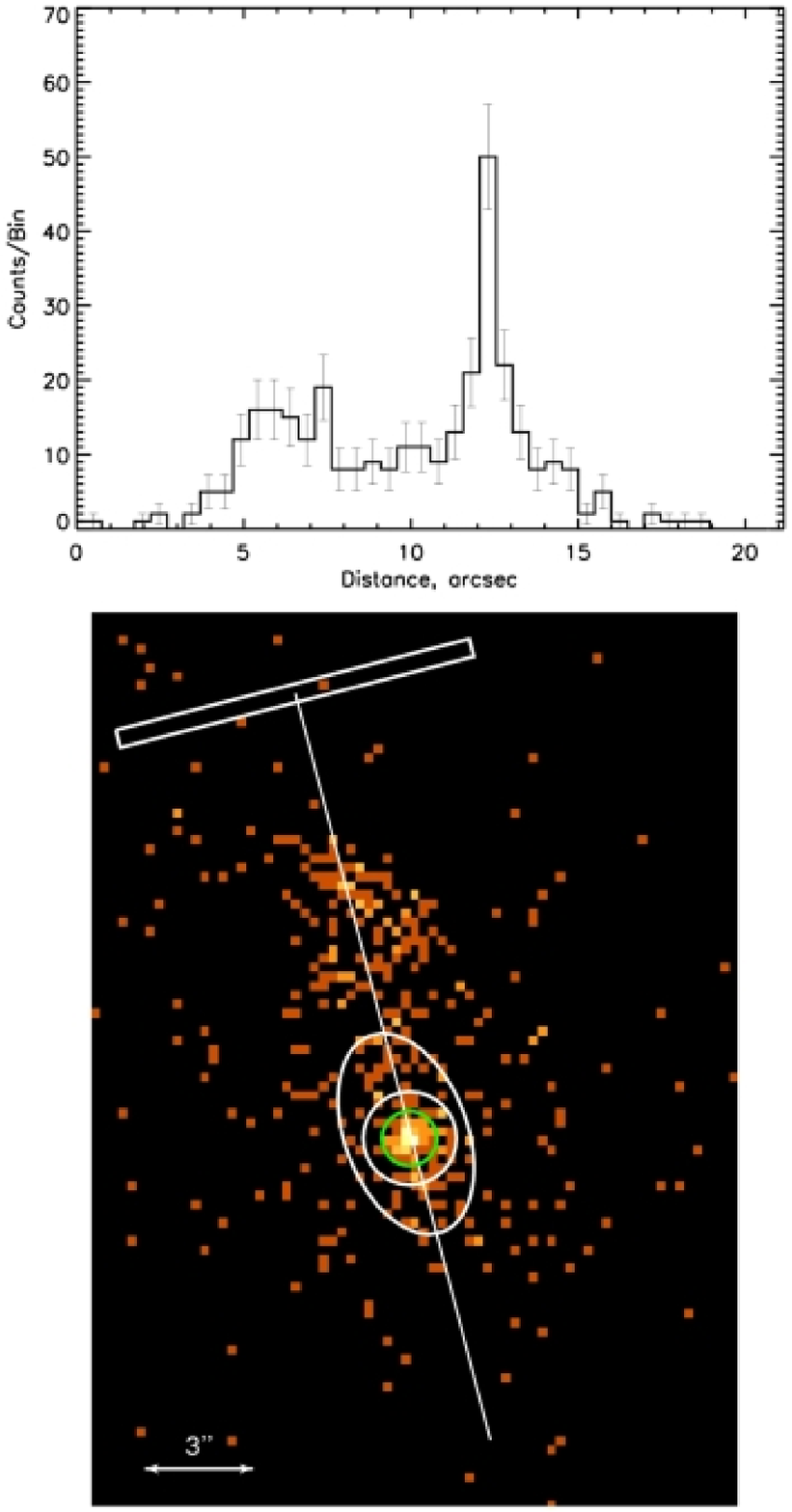}
\caption{{\em Top:}
One-dimensional
surface brightness distribution along the symmetry axis of the J1809 PWN.
{\em Bottom:}
Extraction regions used to measure the pulsar's spectrum
(see \S2.2.2) and the sliding box used for measuring the one-dimensional
surface brightness distribution (see \S2.1 for details).
}
\end{figure}

\subsection{Images}

Figure 1 shows the ACIS-S3 image of the region around J1809. An
extended X-ray source is clearly seen in the image around
R.A.$=18^{\rm h}09^{\rm m}43.123^{\rm s}$, decl.$=-19^{\circ}17'
38.17''$ (these are the coordinates of the center of the brightest
pixel).  The difference of $0.5''$ between this position and the radio
position from Morris et al.\ (2006) is within the uncertainty of
absolute {\sl Chandra} astrometry ($0.6''$ at the 90\% confidence
level).
 The close match between the X-ray and radio positions and
  the extended morphology of the observed
X-ray emission allow us to conclude that we detected the X-ray
emission from J1809 and its PWN.

The brighter {\em inner PWN} component of an $\approx
12''\times3''$ size (i.e. $0.2\times 0.05$ pc$^2$ at $d=3.5$ kpc) is
elongated
 along  the approximate symmetry axis
(position angle $\approx14^{\circ}$ east of north). The linear profile
of the surface brightness distribution along the symmetry axis,
extracted with the $4.9''\times0.49''$ (i.e.\ $10\times1$ pixels) sliding
box, is shown in Figure 2. The inner PWN is surrounded by a similarly
elongated $\sim20''\times40''$  ``halo'' of lower surface brightness,
which we will call the {\em outer PWN} component (see Fig.\ 1, {\em
top}).
 In addition to these two relatively compact
components,
the  heavily binned and smoothed image (see Fig.\ 1, {\em bottom
right}, and  Fig.\ 3{\em a}) reveals
 even fainter large-scale {\em extended emission},
 concentrated at the lower half
 of the S3 chip (south of the pulsar).
The morphology of this emission (see also the {\sl ASCA} image in
Fig.\ 3{\em c}) indicates that it possibly extends further
   south but may not be discernible on the S2 chip because of its lower sensitivity.

\begin{figure*}[t]
 \centering
\includegraphics[width=6.3in,angle=0]{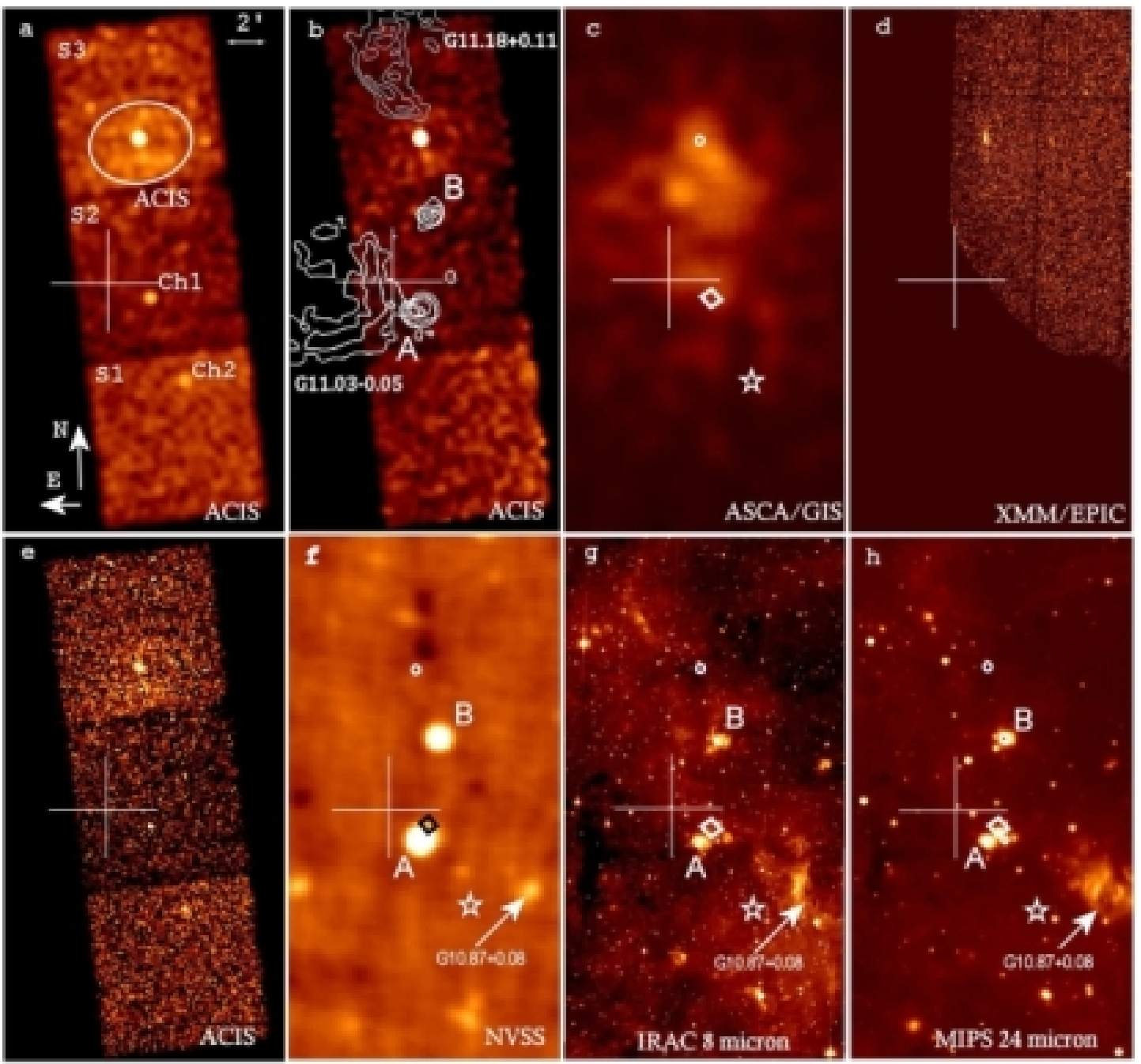}
 \caption{ J1809 and its vicinity at different wavelengths.
All the eight panels show the same area on the sky. {\em a:} ACIS-S3,
S2 and S1 images (0.8$-$7 keV; smoothed with $23.6''$ gaussian
kernel). The white ellipse shows the region used to estimate the flux
of
 the large-scale diffuse emission (see \S2.2.1).
{\em b:}
 The same image divided by the exposure map with point sources
removed (except for J1809). The contours  show the radio emission
from two SNRs and two compact (possibly HII) regions (labeled A and
B; adopted from Brogan et al.\ 2004).
 {\em c:}   {\sl ASCA} GIS2 and GIS3 combined image (39 ks
 total exposure;
0.5$-$10 keV) of the same region (the image has been divided by the
exposure map; astrometry
 has been corrected, courtesy of
E.\ Gotthelf). {\em d:} Combined 9 ks EPIC-PN and MOS1+2 image
(0.5--10 keV) obtained in the Galactic plane survey (PI: R.\ Warwick).
{\em  e:} The same as in the   {\em panel
 a} with no smoothing applied (pixel size is $7.4''$).
{\em f:} NRAO VLA Sky Survey (NVSS; Condon et al.\
 1998)  image at 1.4 GHz.
{\em g:}
 {\sl Spitzer} IRAC 8~$\mu$m image from the GLIMPSE survey.
{\em h:} {\sl Spitzer} MIPS 24~$\mu$m image.
   The white circle in {\em panels c, f, g,} and {\em h} marks
the position of the pulsar, the diamond and the star mark the Ch1 and
Ch2 positions, respectively, and
     the
cross shows the
%
position of the peak of the TeV brightness distribution of HESS\,J1809.
}
\end{figure*}

 The only
source on the S2 chip detected above the $3 \sigma$ level is located
at ${\rm R.A.}=18^{\rm h}09^{\rm m}40.725^{\rm s}$, ${\rm
decl.}=-19^{\circ}25'44.10''$ (the $1\sigma$ centroid uncertainty is
$0.28''$ and $0.29''$  in right ascension
 and declination,
respectively),  well within the brightest central part of HESS~J1809
(about $2'$ west of the peak of the TeV brightness distribution; see
Fig.\ 3).  We designate this source CXOU~J180940.7$-$192544 and call
it Ch1 hereafter. Although Ch1 appears to be extended in the ACIS
image, the point spread function (PSF) simulation shows that this is
likely
 the result of the off-axis location
(off-axis angle $\theta=7\farcm4$). We find no significant large-scale
  non-uniformities in the X-ray background on the S2 chip (see Fig.\ 3).

We have also examined the S1 chip image and found only one source
detected above
 the $3\sigma$ level.
The source, CXOU~180933.3$-$192959 (hereafter Ch2), is located at
${\rm R.A.}=18^{\rm h}09^{\rm m}33.336^{\rm s}$, ${\rm
decl.}=-19^{\circ}29'59.89''$ (the $1\sigma$ centroid uncertainty is
$0.70''$ in R.A. and $0.74''$ in decl.), which is
 about $6'$ from
the  center of HESS~J1809.  The X-ray source is consistent with being
point-like; however, due to the broadened PSF (FWHM $\approx10''$)
at the large off-axis angle ($11.8'$), a compact extended source
cannot be ruled out. We have also searched for diffuse emission
features on the S1 chip and produced an image corrected for the
exposure map non-uniformities (see Fig.\ 3{\em b}). In this image the
X-ray emission is systematically brighter toward the western edge of
the S1 chip, likely due to the imperfection of the mono-energetic (2
keV) exposure map correction at large off-axis angles (see, however,
\S2.5). We found no traces of the radio SNRs G11.03--0.05 and
G11.18+0.11 in the ACIS images (see Fig.\ 3).

\subsection{Spectral analysis}

\subsubsection{PWN spectrum}

We extracted the PWN spectra from two elliptical regions shown in
Figure 1 ({\em top right}). The smaller elliptical region (of 25.1
arcsec$^2$ area) encompasses the brighter inner PWN, while the
larger elliptical region
 (of 399.4 arcsec$^2$ area) includes
the outer PWN component of a lower surface brightness. To avoid the
contamination of the PWN spectrum by the pulsar, we excluded from
these regions the circular region of $1.46''$ radius centered on the
brightest pixel. The background (367 counts in 5202 arcsec$^2$ area,
0.3--8 keV band) was
 measured from the
$37''<r<55''$ annulus centered on the pulsar.
 The total numbers of counts
extracted from the smaller and larger elliptical regions  (excluding the
$1.46''$ radius circle)
 are 153 and
365, of which 99.2\% and 92.4 \% are expected to come from the
source,
 which gives
$151\pm 12$ and $337\pm 19$
 PWN counts in the two regions.
The observed PWN fluxes (in the 1--6 keV band) are $F_{\rm inner} =
(8.1\pm0.7)\times 10^{-14}$ and $F_{\rm pwn}=(14.7\pm0.8) \times
10^{-14}$ ergs s$^{-1}$ cm$^{-2}$ for the inner and the entire
(inner+outer) PWN, respectively. The corresponding average
intensities are $I_{\rm inner}=(4.4\pm 0.4)\times 10^{-15}$ and
$I_{\rm pwn}= (3.7\pm 0.2)\times 10^{-16}$ ergs cm$^{-2}$ s$^{-1}$
arcsec$^{-2}$.

To investigate the PWN spectral properties, we first fit the spectra for
each of the two PWN regions with the absorbed PL model, allowing
the hydrogen column density, $n_{\rm H}$, to vary.
 These fits
result in
 spectral slopes $\Gamma_{\rm pwn}\approx1.2$--2.0
 and
$n_{\rm H,22} \sim0.5$--1.1 (the ranges correspond to the 68\%
confidence level for a single interesting parameter). The difference in
the best-fit parameters for the inner and the entire PWN (see
 Fig.\ 4)
 is statistically insignificant.
 In particular, we see no
spectral softening (expected due to synchrotron cooling)
 in the spectrum extracted from the larger region
 (entire PWN)
 compared
to the spectrum of the inner PWN\footnote{We have also measured
the spectrum of the outer PWN separately and found that the best-fit
PL  parameters are consistent with those obtained for the inner PWN
and the entire PWN spectra, but their uncertainties are larger because
of the larger background contribution.}. Therefore, below we will use
the better constrained best-fit parameters
  for the entire PWN
 (see Figs. 5 and 6) whenever we refer to the PWN spectral properties.

Given the J1809's dispersion measure, ${\rm DM}=197$ cm$^{-3}$ pc
(i.e., electron column density $n_e = 6.08\times 10^{20}$
cm$^{-2}$), the $n_{\rm H,22}$ value of 0.72 (obtained from the fit to the spectrum of the
entire PWN) corresponds to the ISM
ionization degree $n_e/n_{\rm H} \approx 8.4\%$, only slightly below
the commonly used value of 10\%, which corresponds to $n_{\rm H,22}=
0.61$. The total Galactic HI column density in this direction is
$1.8\times 10^{22}$ cm$^{-2}$ (Dickey \& Lockman, 1990).

\begin{figure}
 \centering
\includegraphics[width=2.5in,angle=90]{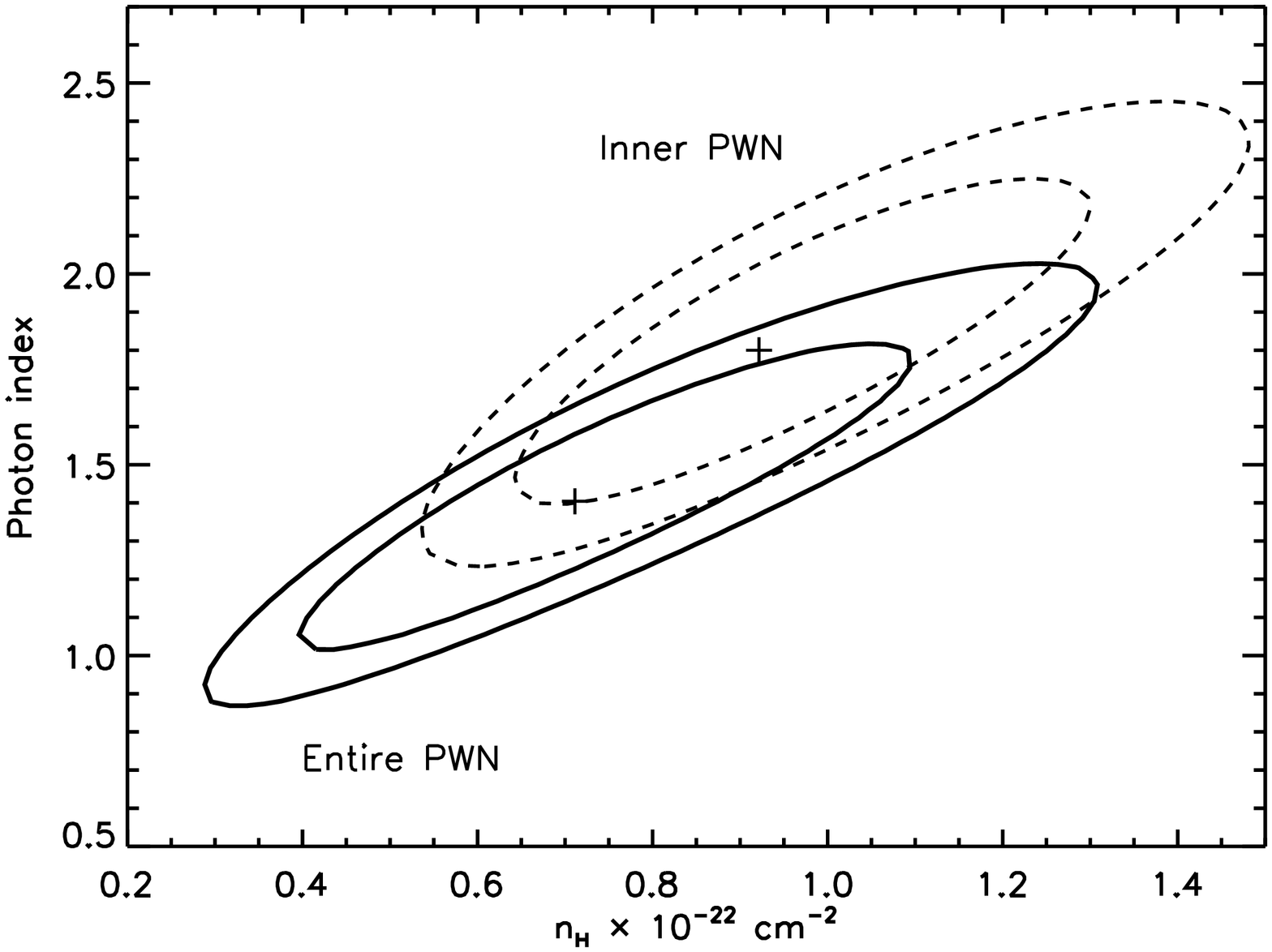}
\caption{Confidence contours (68\% and 90\%) in the $n_{\rm
H}$--$\Gamma$ plane for the PL fit to the inner ({\em dashed})
 and entire ({\em solid}) PWN spectra. The contours
 are obtained with the PL normalization fitted at each point
of the grid.}
\end{figure}

The isotropic luminosity of the entire (inner+outer) X-ray PWN is
$L_{\rm pwn}\equiv 4\pi d^2 F_{\rm pwn}^{\rm unabs}\approx
(3.9\pm0.3)\times 10^{32}d_{3.5}^{2}$ ergs s$^{-1}$ in the 0.5--8 keV
band ($d_{3.5} = d/3.5\, {\rm kpc}$), with approximately equal
contributions from the outer and the inner PWN components (see
Table 1).

 We have also attempted to fit the spectrum
of the faint large-scale emission (within the ellipse shown in Fig.\
3{\em a}, excluding the compact PWN; $13.2$ arcmin$^{2}$ area)
 surrounding J1809 and its compact
PWN. Because of the low surface brightness of this emission and large
background contribution ($\sim65$\%), the spectral fits yield
inconclusive results. With the absorption being fixed at $n_{\rm
H,22}=0.72$, the fits with single PL and thermal plasma\footnote{The
model ``mekal'' in XSPEC, with standard abundance} models give
$\Gamma = 2.3\pm 0.3$ ($\chi_\nu^2 \approx 2$) and $kT=0.67\pm
0.08$ keV, ($\chi_\nu^2 \approx  3$). The main contribution to the
large values of $\chi_\nu^2$ comes from high-energy channels
($\gtrsim 4$ keV and $\gtrsim 2$ keV for PL and mekal, respectively),
which suggests either a mixture of thermal and hard non-thermal
emission or varying $\Gamma$ or $kT$ within the extraction region.
The measured  flux, independent of the
 model, is
$F_{\rm ext}\sim5\times10^{-13}$ ergs s$^{-1}$ cm$^{-2}$ in the
0.8--7 keV band, corresponding to the average surface brightness
$I_{\rm ext}\sim 1\times 10^{-17}$ ergs cm$^{-2}$ s$^{-1}$
arcsec$^{-2}$.
 The unabsorbed
flux, obtained from the PL model, is $F_{\rm ext}^{\rm unabs}\sim
8\times10^{-13}$ ergs s$^{-1}$ cm$^{-2}$ in the 0.5--8 keV band.

\subsubsection{Pulsar spectrum}

To minimize the contamination by the PWN, the pulsar spectrum was
extracted from a small circular aperture (green circle in the bottom
panel of Fig.\ 2) with the radius of 1.5 ACIS pixels ($\simeq 0.74''$,
85\% encircled energy radius), while the background was taken from
the 10 arcsec$^{2}$ region between the white circle and white ellipse
in the bottom panel of Fig.\ 2. The background region includes the
bright part of the PWN; it contributes $\approx
8$ counts to the total of 67 counts extracted from the source
aperture.
 Given the small number of counts
and the large background contribution, we chose not to subtract the
background but rather to fit it simultaneously with the source
spectrum, using an additional absorbed PL model with the same
$n_{\rm H}$ as for the source.
 The pulsar's absorbed flux
is $F_{\rm psr}=(1.8\pm0.2) \times10^{-14}$ ergs cm$^{-2}$ s$^{-1}$ in the
1--6 keV band (aperture corrected).

 Although the
absorbed PL model formally fits the pulsar spectrum indicating a soft
PL ($\Gamma_{\rm psr}\approx2.6$; see Table 2 and Figs.\ 7 and 8),
the fit yields $n_{\rm H,22}\approx 0.4$, smaller than that for the PL
fit to the entire PWN spectrum.
  To obtain a better constrained fit, we
 fixed the hydrogen column density at $n_{\rm
H,22}=0.72$, obtained above from the PL fit to the PWN spectrum.
With this $n_{\rm H}$, the single PL fit is still acceptable,
 but it yields a
large photon index, $\Gamma=3.2\pm0.4$, suggesting a thermal
emission contribution. A two-component, BB+PL, fit yields reasonable
values of fitting parameters (see Table 2 and Figs.\ 7--9), which,
however, are poorly constrained because of the small number of
photons detected. The slope of the PL component is $\Gamma_{\rm
psr}= 1.2\pm 0.6$, and its unabsorbed luminosity is $L_{\rm
psr}\sim4 \times 10^{31}d_{3.5}^2$ ergs s$^{-1}$ in the 0.5$-$8 keV
band. The temperature and the projected area
 of the BB component are
strongly correlated (see Fig.\ 9), which results in large
uncertainties for these parameters. The
 best-fit temperature for the BB component
is $T\approx2$ MK,
  while the projected emitting
  area,
 $\mathcal{A}\sim 3 \times 10^{6}d_{3.5}^2$ m$^2$,
is smaller than  that of the surface of a neutron star ($\pi
R^{2}\sim3\times 10^{8}$ m$^{2}$), but
 larger than the
conventional polar cap area $A_{\rm pc}=2\pi^2R^3/cP\approx
2.5\times10^5$
 m$^{2}$. The corresponding bolometric luminosity,
$L_{\rm bol} \equiv 4{\mathcal{A}}\sigma T^4 \sim 1\times 10^{32}d_{3.5}^2$
ergs s$^{-1}$.

\begin{figure}
 \centering
\includegraphics[width=3.2in,angle=0]{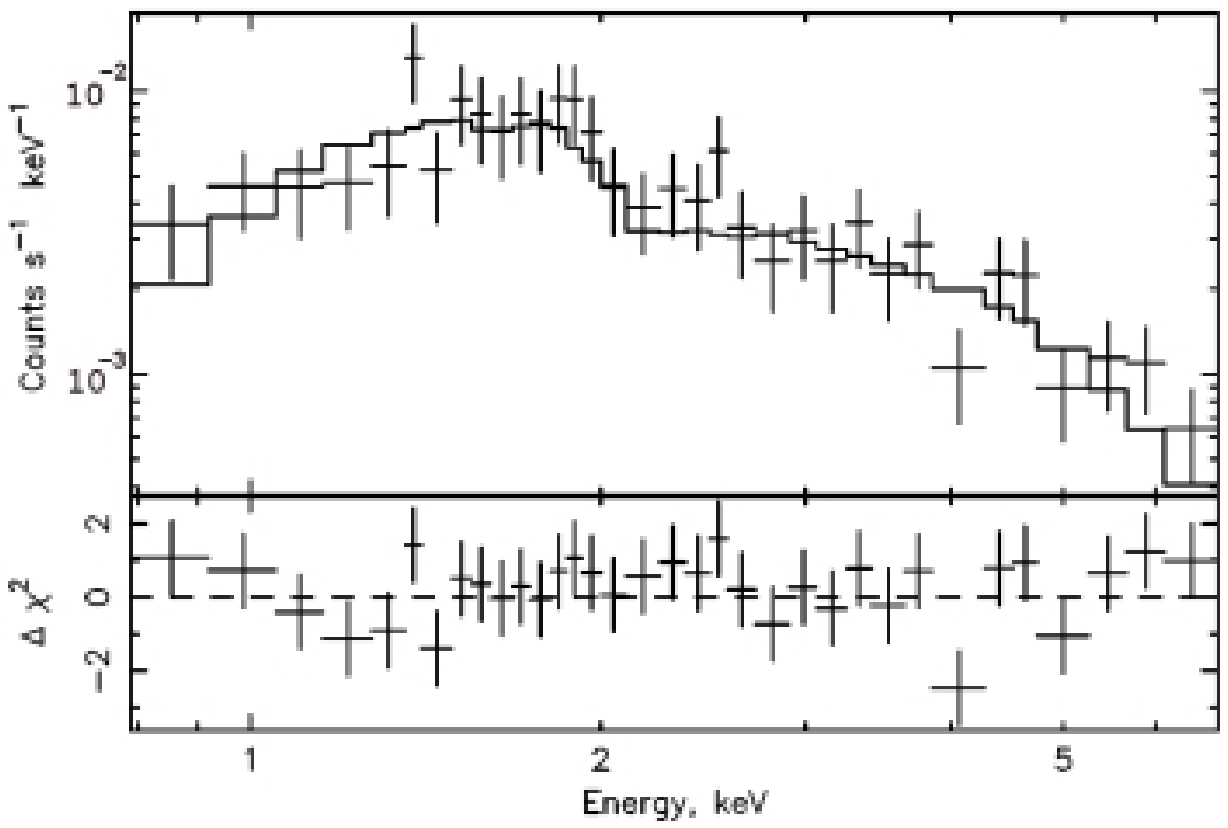}
 \caption{  Entire (outer+inner)
PWN spectrum fitted with the PL model (see Table 1 for details).
}
\end{figure}

\begin{figure}
 \centering
\includegraphics[width=2.5in,angle=90]{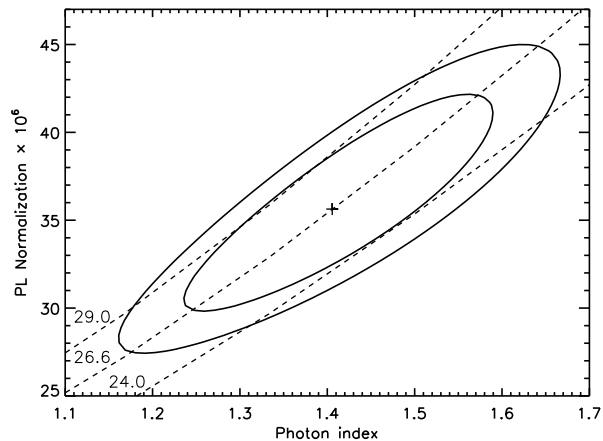}
\caption{  Confidence contours (68\% and 90\%) for the PL fit to the
 entire PWN spectrum (for a fixed $n_{\rm H,22}=0.72$).
 The PL normalization
is in units of $10^{-6}$ photons cm$^{-2}$ s$^{-1}$ keV$^{-1}$ at 1
keV. The dashed curves are the loci
 of constant unabsorbed flux in the 0.5--8 keV band;
the flux values near the curves are in units of 10$^{-14}$ ergs
cm$^{-2}$ s$^{-1}$.
 }
\end{figure}

\begin{figure}
 \centering
 \vbox{
\includegraphics[width=3.2in,angle=0]{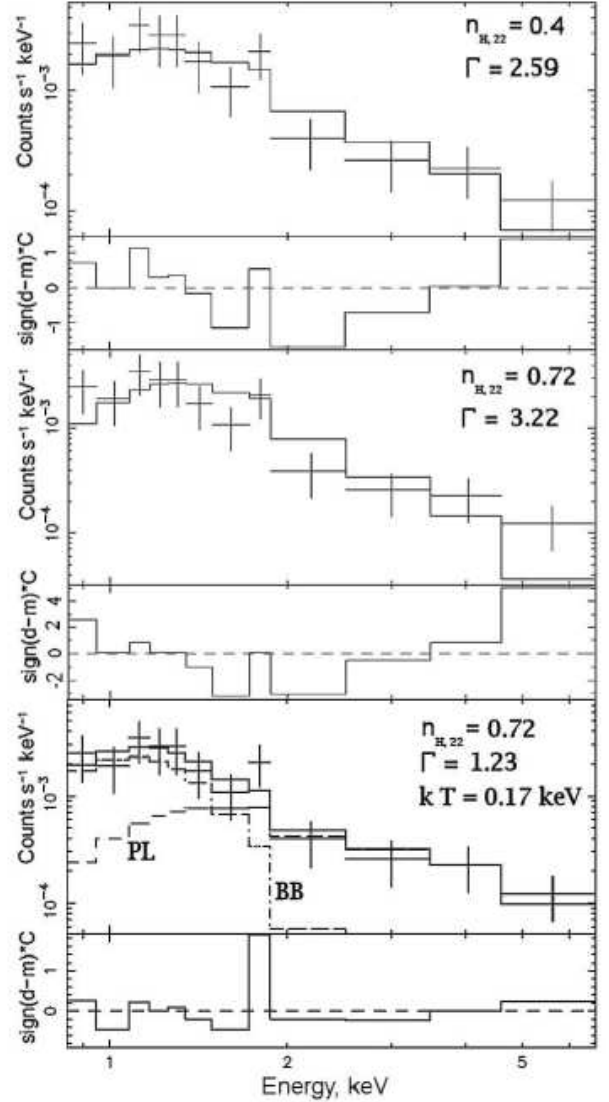}
} \caption{ {\em Top:} Pulsar  spectrum fitted with the absorbed PL
model. {\em Middle:} Pulsar spectrum fitted with the absorbed PL model
where the $n_{\rm H}$ was fixed at the best-fit value for the  entire
PWN
($n_{\rm H,22}=0.72$). {\em Bottom:} The pulsar spectrum fitted with the
PL+BB model with
fixed $n_{\rm H,22}=0.72$. The dashed and dash-dotted histograms
correspond to
the BB and PL components, respectively.
The residual panels show the contributions of the energy bins into
the best-fit C-statistic (multiplied by $-1$ when the number of data
counts is smaller than the number of model counts).
 }
\end{figure}

\begin{figure}
  \centering
 \vbox{
\includegraphics[width=3.2in,angle=0]{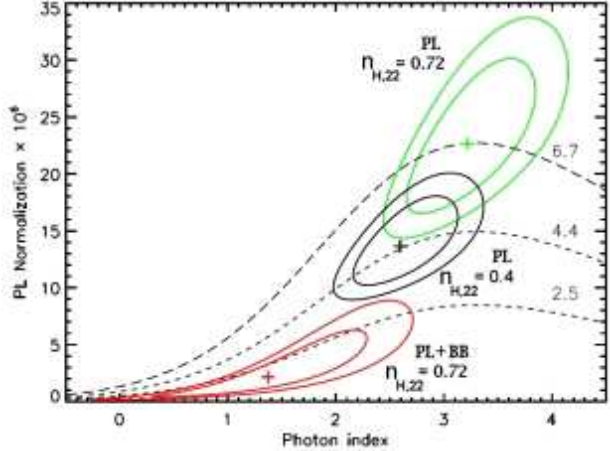}
} \caption{ Confidence contours (68\% and 90\%) for the PL fit to the pulsars's spectrum with
$n_{\rm H,22}=0.72$ (green), PL+BB fit with $n_{\rm H,22}=0.72$
(red)
 and PL fit with
$n_{\rm H,22}=0.4$ (black) which is the best-fit value for the PL
model.
 The PL normalization
is in units of $10^{-6}$ photons cm$^{-2}$ s$^{-1}$ keV$^{-1}$ at 1
keV. The dashed curves are the lines of constant unabsorbed flux
in the 0.5--8 keV band (the flux values are in units of $10^{-14}$
ergs cm$^{-2}$ s$^{-1}$).
}
\end{figure}

\begin{figure}[]
 \centering
\includegraphics[width=3.2in,angle=0]{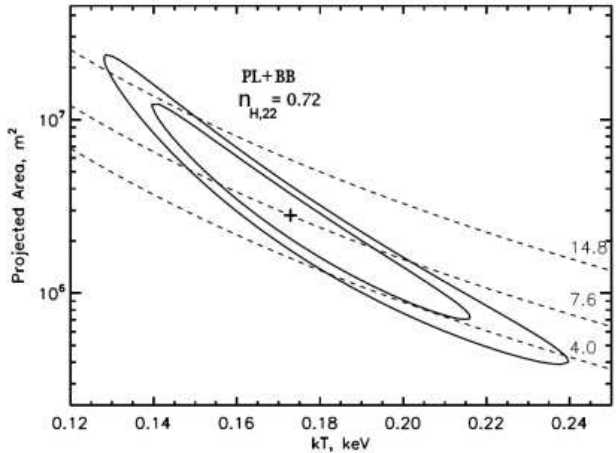}
\caption{ Confidence contours (68\% and 90\%) for the BB component
of the BB+PL fit to the pulsar's spectrum, for $n_{\rm H,22}=0.72$.
 The BB normalization (vertical axis) is the
projected emitting area in units of m$^2$, assuming the distance of
3.5 kpc. The lines of constant bolometric flux (in units of
10$^{-14}$ ergs cm$^{-2}$ s$^{-1}$),
 are plotted as dashed lines. }
\end{figure}

\begin{table}[]
\caption[]{PL fits to the PWN spectrum} \vspace{-0.5cm}
\begin{center}
\begin{tabular}{ccccccc}
\tableline\tableline Model & $n_{\rm H,22}$   &
$\mathcal{N}$\tablenotemark{a}  &
$\Gamma$ & ($C$ or $\chi^2$)\tablenotemark{b}/dof   & $L_{\rm X}$\tablenotemark{c} \\
\tableline

  Entire PWN           &       $0.72$     &
$35.6^{+4.6}_{-3.8}$       &
$1.41^{+0.12}_{-0.11}$          &  $0.82/34$  & $3.89_{-0.27}^{+0.24} $  \\
  Inner    PWN       &       $0.98$                     &
$30.6^{+5.3}_{-4.6}$ &       $1.85_{-0.17}^{+0.17}$   &  $335/526$  & $2.21_{-0.17}^{+0.19} $  \\
      \tableline
\end{tabular}
\end{center}
\tablecomments{The fits are for fixed $n_{\rm H,22}\equiv n_{\rm
H}/10^{22}$ cm$^{-2}$. The uncertainties are given at 68\%
confidence level for a single interesting parameter. }
\tablenotetext{a}{Spectral flux in units of $10^{-6}$ photons
cm$^{-2}$ s$^{-1}$ keV$^{-1}$ at 1 keV.} \tablenotetext{b}{We use the
C statistic (Cash 1979) for the inner PWN (which has less counts) and
the $\chi^2$ statistic for the entire PWN.}
\tablenotetext{c}{Unabsorbed isotropic luminosity in the 0.5--8 keV
band, in units of $10^{32}$ ergs s$^{-1}$. }
\end{table}

\begin{table}[]
\caption[]{Fits to the pulsar spectrum} \vspace{-0.5cm}
\begin{center}
\begin{tabular}{ccccccc}
\tableline\tableline Model & $n_{\rm H,22}$   &
$\mathcal{N}$\tablenotemark{a} or $\mathcal{A}$\tablenotemark{b} &
$\Gamma$ or $kT$\tablenotemark{c} & $C$  & $L_{\rm X}$\tablenotemark{d} or $L_{\rm bol}$\tablenotemark{e} \\
\tableline

PL & $0.72$ & $22.6_{-4.2}^{+4.9}$ & $3.22_{-0.38}^{+0.42}$ & $412$ & $0.98_{-0.17}^{+0.22}$  \\
PL & $0.4$ & $13.6^{+3.1}_{-2.4}$ & $2.59_{-0.28}^{+0.34}$   &  $411$    & $0.65_{-0.09}^{+0.10}$  \\
PL+BB(PL)       &     $0.72$ & $2.84^{+2.66}_{-1.51}$ &
$1.23\pm0.62$           &  $407$   & $0.37^{+0.12}_{-0.10}$  \\
  PL+BB(BB)       &  $0.72$ & $2.8^{+5.6}_{-1.8}$ &
$0.17_{-0.03}^{+0.03}$   &  $407$   &
$1.0^{+0.6}_{-0.4}$   \\
  \tableline
\end{tabular}
\end{center}
\tablecomments{The fits are for fixed $n_{\rm H,22}\equiv n_{\rm
H}/10^{22}$ cm$^{-2}$. The uncertainties are given at 68\%
confidence level for a single interesting parameter. In each case the
fits were done using the C statistics (Cash 1979) and the unbinned
source and background spectra, with the total of 1052 channels. }
 \tablenotetext{a}{Spectral flux in units of $10^{-6}$
photons cm$^{-2}$ s$^{-1}$ keV$^{-1}$ at 1 keV.}
\tablenotetext{b}{Projected area of the emitting region for the BB
model, in units of $10^{6}$ m$^2$.} \tablenotetext{c}{BB temperature
in keV.} \tablenotetext{d}{Unabsorbed PL luminosity in the 0.5--8 keV
band, in units of $10^{32}$ ergs s$^{-1}$.}
\tablenotetext{e}{Bolometric BB luminosity, in units of $10^{32}$ ergs
s$^{-1}$.}

\end{table}

\subsubsection{ Ch1 spectrum}

  We extracted 99 events from the elliptical
 region (with the semi-major and semi-minor  axes of $10.3''$ and $6.3''$,
respectively)
centered on the best-fit
  position of Ch1 and
  conforming to the shape of the off-axis PSF.
Based on the PSF simulation, the extraction region contains
 $\approx95\%$ of  point source counts. The expected background contribution, 4.4 counts,
 is negligible ($\approx4\%$ of the total number of counts).
  The measured source flux
  in
 0.8--7 keV is $F_{\rm Ch1}=(8.6\pm0.9)\times10^{-14}$
   ergs cm$^{-2}$ s$^{-1}$, after correcting
    for vignetting and finite extraction aperture size.
The spectrum fits well by an absorbed PL model (see Table 3) with
 $n_{\rm H,22}=1.2\pm0.4$ and $\Gamma=1.4\pm0.4$.
The unabsorbed flux in 0.5--8 keV is
 $F_{\rm Ch1}^{\rm unabs}\approx1.4\times10^{-13}$ ergs cm$^{-2}$ s$^{-1}$.

\subsubsection{ Ch2 spectrum}

   We extracted 90 events within a $30''$ radius around
the best-fit
  position of Ch2. Based on the PSF simulation, the extraction region contains
 $\sim85\%$ of the source counts if Ch2 is a point source.
  The background contributes $\approx32$ counts,
 (i.e.\ $\approx36\%$ of the total number of counts). The measured
0.8--7 keV source flux is $F_{\rm Ch2}\approx 9\times10^{-14}$
   ergs cm$^{-2}$ s$^{-1}$, after correcting
    for vignetting and finite extraction aperture size
(assuming a point source). The thermal plasma (mekal)
 model fits the Ch2 spectrum best, although fits with absorbed
 PL and BB models are also acceptable (see Table 3 for details).

\begin{table}[]
\caption[]{Fits to the Ch1 and Ch2 spectra} \vspace{-0.5cm}
\begin{center}
\setlength{\tabcolsep}{1pt}
\begin{tabular}{cccccccc}
\tableline\tableline Source & Model & $n_{\rm H,22}$   &
$\mathcal{N}$\tablenotemark{a} or $\mathcal{A}$\tablenotemark{b} &
$\Gamma$ or $kT$\tablenotemark{c} & ($C$ or $\chi^2$)\tablenotemark{d}/dof
& $F_{\rm X}^{\rm un}$ or $F_{\rm bol}$\tablenotemark{e} \\
\tableline

Ch1 & PL & $1.17^{+0.43}_{-0.39}$ & $17.8_{-7.3}^{+12.8}$ & $1.42_{-0.38}^{+0.40}$ & $291/523$ & $1.4\pm0.3$  \\
Ch2 & PL & $0.66 (<1.4)$ & $\sim45$ & $\sim6$ & $2.3/3$ & $\approx4.7$  \\
Ch2 &  BB       &     $0.3 (<1.1)$ & $\sim0.28d_1^2$ & $\approx0.16$ &
$2.4/3$   & $\approx0.90$ \\
 Ch2 &  mekal       &
$0.6_{-0.2}^{+0.6}$ & $\sim2.6$ & $0.7\pm0.2$ & $1.1/3$   &
$\approx0.54$

   \\
  \tableline
\end{tabular}
\end{center}
\tablecomments{ The uncertainties are given at the 68\% confidence
level for a single interesting parameter. For Ch2, the upper limits on
$n_{\rm H}$ value (at 68\% confidence) are given in brackets, the
lower limits are not constrained at the same confidence level. }
\tablenotetext{a}{Normalization for the PL model is the spectral flux
at 1 keV in $10^{-6}$ photons cm$^{-2}$ s$^{-1}$ keV$^{-1}$.
Normalization for the mekal model is the Emission Measure (EM) in
$10^{53}$ cm$^{-3}$, scaled to a distance of 1 kpc. }
\tablenotetext{b}{Projected area of the emitting region for the BB
model, in units of $10^{6}$ m$^2$, normalized to $d=1$ kpc.}
\tablenotetext{c}{BB temperature in keV.} \tablenotetext{b}{We use
the C statistic for Ch1 and the $\chi^2$ statistic for Ch2.}
\tablenotetext{e}{Unabsorbed PL flux in the 0.5--8 keV band or
bolometric BB flux, in units of $10^{-13}$ ergs  cm$^{-2}$ s$^{-1}$
(corrected for vignetting and finite extraction aperture size).}
\end{table}

\subsection{Timing}

The 3.24 s time-resolution of this observation is insufficient to
observe the 82.7 ms pulsations from the pulsar. We found no
significant variability in the Ch1 and Ch2 lightcurves.

\subsection{Archival X-ray data}

The region of interest has been previously observed by {\sl ASCA} for
39 ks
 (Bamba et al.\ 2003). The {\sl ASCA} GIS image in Figure 3{\em c} shows
 a region of enhanced  X-ray brightness
 that encompasses J1809 and
its compact PWN; however, the diffuse X-ray emission also extends at
least $\sim10'$ southward,
 covering the
central region of
 the brightest part of HESS~J1809.

  The field was also partly observed
 by the {\sl XMM-Newton} as a part of the Galactic plane survey (PI: R.\ Warwick). A point-like object is clearly
 seen at the pulsar position in  the combined EPIC
(MOS1+MOS2+PN) image shown in Figure 3.
 However, the short
 8 ks exposure, the off-axis location, and the high EPIC background
do not allow one to detect the faint
 extended component, while the compact PWN cannot be resolved from the pulsar because of the  broad
  PSF of {\sl XMM-Newton}.

\subsection{Optical-IR-radio data}

To understand the nature of the X-ray sources Ch1 and Ch2, projected
within the HESS source image and look for other sources that could be
related to the TeV emission, we have examined the field at other
wavelengths. We found no counterparts to Ch1
 in the Two Micron All Sky Survey
(2MASS; Skrutskie et al.\ 2006) or Digital Sky Survey (DSS2)\footnote{
see http://archive.eso.org/dss/dss}
 catalogs,
up to the limiting magnitudes $K_s=15.4$, $H=16$, $J=17.5$, $R=19$,
and $B=21$.  The nearest optical/NIR
 source is a 2MASS  point source
($J=15.45\pm0.05$, $H=13.10\pm0.05$,
	$K=12.06\pm0.04$) located at ${\rm R.A.}=18^{\rm h}09^{\rm
m}40.92^{\rm s}$, ${\rm decl.}=-19^{\circ}25'45.7''$. The $3.2''$ offset
from the best-fit Ch1 position substantially exceeds the position
uncertainty of $\approx 0.4''$. Figure 3 shows that Ch1 is projected
very close to the extended Source A seen in the radio and IR images.

The only NIR/optical source within the $5''$ radius of the Ch2 position
is a star, NOMAD1~0705$-$0568334 in the NOMAD catalog (Zacharias
et al.\ 2005), offset by only $0.9''$ from Ch2. Given the relatively large
uncertainty of the Ch2 position (see \S2.1), the positions of the star
and Ch2 can be considered coincident. The X-ray-to-optical flux ratio,
$F_X/F_V\sim 2\times 10^{-3}$, is typical for a K star with coronal
X-ray emission (Maccacaro et al.\
 1988).
The soft X-ray spectrum of Ch2 and the optical-NIR magnitudes of the
star ($B=14.96$, $V=14.66$, $R=14.19$, $J=13.35$, $H=12.84$, and
$K=12.72$) support such an interpretation. No Ch2 counterpart is
seen in the NVSS 20 cm image or in the {\sl Spitzer} 8 and 24 $\mu$m
images.

The NVSS 20 cm image ($45''$ restoring beam size) and the $25''$
resolution images by Brogan et al.\ (2004)
 show two bright
compact radio sources (marked A and B in Fig.\ 3) projected near the
HESS source center,
 with the 1.4 GHz spectral fluxes
of 0.3 Jy and 0.15 Jy for Source A and Source B, respectively. Within
these sources, 6.7 GHz methanol masers have been detected
(Pestalozzi et al.\ 2005):
  G10.95+0.02 in Source A
(peak flux 15 Jy at 6.7 GHz, $V_{\rm LSR}=24\pm 1$ km s$^{-1}$,
near-distance 3.2 kpc) and G11.03+0.06 in Source B (peak flux 0.7 Jy
at 6.7 GHz, $V_{\rm LSR}=20\pm 1$ km s$^{-1}$, near-distance 2.9
kpc).

Both A and B are also very bright far-IR sources, IRAS\,18067--1927
and IRAS\,18067--1921 in the IRAS Point Source Catalog v2.1 (IPAC
Infrared Science Archive\footnote{http://irsa.ipac.caltech.edu/}), with
peak fluxes of 1292 and 513 Jy at 60 $\mu$m, respectively. The {\sl
Spitzer\/} IRAC 8\,$\mu$m and MIPS 24\,$\mu$m images (Fig.\ 3{\em
g,h}; taken from the GLIMPSE
survey\footnote{http://www.astro.wisc.edu/sirtf/}) show extended
sources of irregular shape at these positions, with characteristic sizes
of $\sim 1'$. They are not seen in the DSS and 2MASS images,
 which suggests that they are intrinsically very cold
($T\sim 70$ K) and/or strongly absorbed molecular/dust complexes.

  The {\sl Spitzer} images also reveal an extended
 source with
an interesting morphology (shell-like in the 24 $\mu$m image),
   located $\sim10'$ southwest
  of the
HESS~J1809 center (see Fig.\ 3{\em g,h}). The source has an extended
radio counterpart (NVSS~180919--192904)
  clearly seen in the 20 cm image shown in the same figure.
It has been detected by Helfand et al.\ (2006) in the Multi-Array
Galactic Plane Survey (MAGPIS) and proposed to be an SNR candidate
G10.8750+0.0875.  It is possible that this extended source contributes
to the non-uniform X-ray background on the S1 chip (see \S2.1 and
Fig.\ 3{\em b}). The SNR candidates G11.03$-$0.05 and G11.18+0.11
  (Brogan et al.\
  2004; see Fig.\ 3{\em b})
   are too faint to be seen in the shallow NVSS images.

\section{Discussion.}

The sub-arcsecond angular resolution of {\sl Chandra} has allowed us
to resolve the compact J1809 PWN, disentangle the pulsar and the
extended emission components, and measure their properties
separately. Thanks to the very low ACIS background, we were also
able to detect the faint large-scale emission surrounding the compact
PWN. The  observation has also provided serendipitous coverage of
the central part of the extended TeV source
  HESS\,J1809 and revealed
two X-ray sources bright enough to permit  spectral measurements and
 accurate determination of their positions.
In this section we discuss some implementations of our findings.

\subsection{The J1809 PWN}

\subsubsection{Luminosity and spectrum}

At the plausible distance of 3.5 kpc, the
unabsorbed
X-ray luminosity
 of the
compact PWN, $L_{\rm pwn}\sim 4 \times 10^{32}$ ergs s$^{-1}$,
corresponds to the X-ray efficiency, $\eta_{\rm pwn}\equiv L_{\rm
pwn}/\dot{E} \sim 2.2\times10^{-4}$, similar to
 those of PWNe around younger and more powerful Vela-like pulsars
(see Fig.\ 10).  Some of the detected
 X-ray PWNe associated with older and less powerful pulsars
 have similar or higher X-ray efficiencies (e.g.,  $\eta_{\rm pwn}\sim 2\times 10^{-3}$
and $4\times 10^{-4}$  for  J1509--5850 [$\tau=150$ kyr] and
B0355+54 [$\tau=560$ kyr], respectively; Kargaltsev et al., in
preparation), while others show much lower efficiencies (e.g.,
$\eta_{\rm pwn} \sim 5\times 10^{-5}$ and  $7\times 10^{-6}$ for PRS
J1740+1000 [$\tau=100$ kyr] and Geminga [$\tau=340$ kyr],
respectively; Kargaltsev et al., in preparation; Pavlov et al.\ 2006).
 This indicates that
$\eta_{\rm pwn}$ is not significantly correlated with $\tau$ or
$\dot{E}$, at least for young and middle-aged pulsars ($\tau \lesssim
1$ Myr), perhaps because, in addition to $\dot{E}$ and $\tau$, it
depends on other factors (e.g., the pulsar's speed and the angle
between the spin and magnetic axes).

The spectral slope of the J1809 PWN, $\Gamma_{\rm pwn} =1.4\pm
0.1$,
 is similar to those of the PWNe
around Vela-like pulsars (listed in
 Table 2 of Kargaltsev et al.\
2007a, hereafter KPG07a), except for
two bow-shock PWNe with prominent tails,
 J1747--2958 (the Mouse; Gaensler
et al.\ 2004)  and B1757--24 (the Duck; Kaspi et al.\ 2001), which show
softer spectra ($\Gamma=2.0\pm0.2$ and $2.5\pm0.3$,
respectively). Comparing the J1809 spectrum with those of older
PWNe, we see that it is similar to the spectra of relatively compact
PWNe (e.g., $\Gamma=1.4\pm0.3$ for the bright part of the
B0355+54 PWN; McGowan et al.\ 2006), but it is harder than the
spectra of the extended tails in
 J1740+1000
 and J1509--5850 PWNe ($\Gamma=1.8\pm0.4$ and $2.2\pm0.3$,
respectively; Kargaltsev et al., in preparation).
 It hints that PWN
spectra are correlated with PWN morphology rather than with the
pulsar age and spindown power.

\begin{figure}[]
 \centering
\includegraphics[width=2.7in,angle=90]{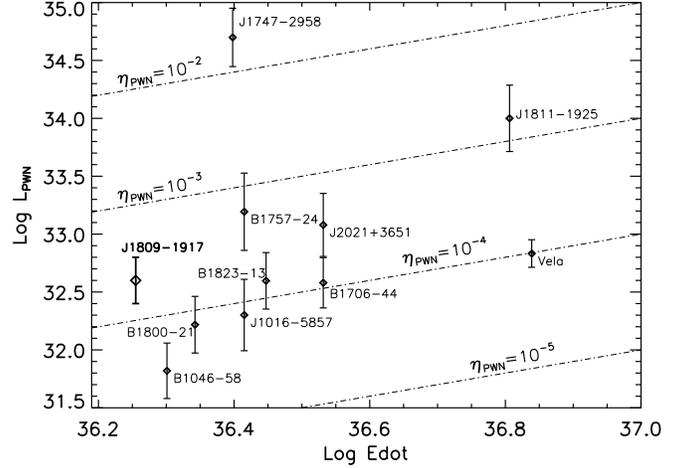}
\caption{ PWN luminosity versus pulsar spin-down power showing
J1809 and 10 Vela-like pulsars  observed with {\sl Chandra}. The
luminosities are estimated for the 0.5--8 keV band. The dash-dot lines
are the lines of constant PWN efficiency, $\eta_{\rm pwn}$. The error
bars include the statistical uncertainties and the nominal 30\%
distance uncertainties, except for the Vela pulsar whose parallax has
been measured (Dodson et al.\ 2003a).
}
\end{figure}

\subsubsection{PWN morphology}

The unknown proper motion of J1809 and the faintness
of the surrounding large-scale emission
 complicate the interpretation of the observed PWN,
but its ``cometary'' appearance strongly suggests that the pulsar's
motion plays a major role.

\medskip
\noindent
{\sl 3.1.2.1. Bow shock in a nearly isotropic pulsar wind? ---}
The compact J1809 PWN is elongated approximately along the
north-south direction, with the pulsar located much closer to its
southern end (see Fig.\ 1).
 Such a
cometary morphology can be attributed to a bow shock created by
the pulsar moving supersonically in the southern direction (${\rm
P.A.}\approx 194^\circ$). The apex of the termination shock (TS) of
the pulsar wind is located at the
 distance
\be R_h \approx \left[\frac{\dot{E} f_\Omega}{4\pi c (p_{\rm
amb}+p_{\rm ram})}\right]^{1/2} \ee
 ahead of the pulsar. At this
distance, the pulsar wind pressure, $p_{w} = \dot{E} f_\Omega (4\pi c
r_s^2)^{-1}$ ($f_\Omega$ takes into account anisotropy of the pulsar
wind), is balanced by the sum of the ambient pressure, $p_{\rm
amb}= \rho kT (\mu m_{\rm H})^{-1}= 1.38\times 10^{-12} n \mu^{-1}
T_4$ ergs cm$^{-3}$, and the ram pressure, $p_{\rm ram}= \rho v^2 =
1.67\times 10^{-10} n v_7^2$ ergs cm$^{-3}$ ($T_4 = T/10^4\,{\rm
K}$, $v_7 = v/10^7\,{\rm cm\, s}^{-1}$, $\mu$ is the molecular weight,
and $n = \rho/m_{\rm H}$ is in units of cm$^{-3}$). Assuming
$p_{\rm ram}\gg p_{\rm amb}$ (or ${\mathcal M} \gg 1$, where
${\mathcal M}=v/c_s$ is the Mach number,
 $c_s=(5 kT/3 \mu m_{\rm H})^{1/2}=
12 \mu^{-1/2} T_4^{1/2}$ km s$^{-1}$
 is the sound speed in the ambient medium),
we obtain $R_{h}= 1.7\times 10^{17}n^{-1/2} f_\Omega^{1/2}
v_7^{-1}$ cm.

The shocked pulsar wind, observed in X-rays as a bow-shock PWN,
 is confined between the TS and the contact discontinuity
(CD) surface. For ${\mathcal M}\gg 1$ and a {\em nearly isotropic}
preshock wind with a low magnetization parameter $\sigma$, the TS
acquires a bullet-like shape (Bucciantini et al.\ 2005, hereafter B05).
The distance $R_h$ between the pulsar and the bullet head is given
by equation (1), while the CD surface head is at a distance $\approx
1.3 R_h$ from the pulsar. Assuming that the sharp rise in the
brightness profile (Fig.\ 2) at about $3''$  from the
 pulsar (i.e.\
$1.6\times 10^{17} d_{3.5}$ cm in the plane of the sky) corresponds
to the CD head,
 we can estimate
the pulsar velocity,
 \be v \sim
140\, n^{-1/2} d_{3.5}^{-1} f_\Omega^{1/2} \sin i\,\,\, {\rm km\,\, s}^{-1},
\ee
and the Mach number,
${\mathcal M}\sim 12 n^{-1/2} T_4^{-1/2} \mu^{1/2} f_\Omega^{1/2} d_{3.5}^{-1}\sin i$,
where $i$ is the angle between the velocity vector and the line of sight.

The observed projected length, $\sim 12''$, and width, $\sim 3''$, of
the inner PWN are close to the length $\sim 7 R_h\sim 15''$, and
width, $\sim R_h\sim 2''$, of the TS bullet,
 predicted by the
B05 models for $\sigma\sim {\rm a\,\,\, few}\times 10^{-3}$. The
width of the outer PWN, $\sim 20''$, can be interpreted as the
diameter of the CD shell (approximately cylindrical behind the pulsar,
with a diameter $\sim 8 R_h \sim 18''$, according to B05).

The B05 models also predict a tail behind the back surface of the TS
bullet, where the collimated shocked pulsar wind flows with
subrelativistic velocities: $0.1c$--$0.3c$ in the inner channel with a
cylindrical radius $\sim R_h$ ($\sim 2''$ in our case), and up to
$0.8c$--$0.9c$ in the outer channel, confined between the cylindrical
surfaces with radii $\sim R_h$ and $\sim 4 R_h$. The lack of an
extended tail in the ACIS images is puzzling; it could be due to a
relatively low Mach number of the pulsar and insufficient sensitivity
of the short ACIS exposure.
 On the other hand, the shape of the faint
thin feature northeast of inner PWN hints that it may be a bent
extension of the inner PWN, which is difficult to explain if this is a
strongly collimated, mildly relativistic flow just behind the back
surface of the TS bullet (unless the flow is subject to kink instabilities).
However, statistical significance of this feature is too low to draw any
definitive conclusions.

\medskip\noindent
{\sl 3.1.2.2. Effects of pulsar wind anisotropy.} --- We should note that
the B05 models assume an isotropic pulsar wind ($f_\Omega =1$).
We know from observations of young pulsars, such as Crab and Vela,
that the wind is not isotropic, but it is mostly confined
to the equatorial plane. In addition, in young PWNe we
often see jets along the pulsar's spin axis. These jets can be formed by
polar outflows originating in the pulsar magnetosphere or they can be
created by tangential inflows just outside the equatorial TS surface
converging toward the spin axis
(Komissarov \& Lyubarsky 2004). We are unaware of PWN models
that include both the pulsar wind anisotropy and the ram pressure
effects. We can, however, expect that, in the case of an anisotropic
wind, the PWN morphology, at least close to the pulsar, would
depend on the orientation of the spin axis with respect to the pulsar's
velocity.

As {\sl Chandra} observations have shown, the spin axis is oriented
along the direction of pulsar's motion in a number of
young pulsars (e.g., Ng \& Romani 2004). In this case, we expect that
 the equatorial
outflow would form a shell between the TS and the CD surface behind
the pulsar, filled by a relativistic plasma with a subrelativistic bulk flow
velocity. The morphology of the X-ray emission from such a PWN
would be generally similar to that in the case of isotropic wind,
although we may expect to see a shell-like (rather than filled) PWN
appearance at sufficiently high resolution. Such a structure (``outer
tails'') is possibly seen in the Geminga PWN (Caraveo et al.\ 2003;
Pavlov et al.\ 2006). As for the jets coaligned with the pulsar velocity
vector, the rear jet is expected to be seen along the axis of the shell
filled by the shocked equatorial wind
 behind the pulsar (at least if the jet originates
from the pulsar magnetosphere), as observed in the Geminga PWN
(Pavlov et al.\ 2006).  The front jet is expected to
pierce the head of the shell (as $f_\Omega \gg 1$ in the jet),
and it can be seen ahead of the pulsar unless
it is crushed by the ram pressure.
The jets could be of quite different brightness and length because
of the Doppler
boosting and different effective pressures in front and behind the pulsar.

In the case of J1809, the high-resolution images shown in Figure 1
   indicate the presence of a narrow structure
(with a linear extent of $\simeq 3''$)
  just
  south of the pulsar, which might be interpreted as a front jet.
Furthermore, the heavily binned low-resolution image (Fig.\ 3{\em b})
shows a faint structure extending in the direction of the presumed
pulsar's proper motion. Although  tentative, the structure is
consistent
   with the X-ray morphology
seen in the {\sl ASCA} GIS image (Fig.\ 3{\em c}).
  Therefore, one could speculate that
these structures are connected to each other and could be parts of the front
 jet.
In this interpretation, the rear jet may contribute to the inner PWN
emission, and the above-mentioned bent extension of the inner PWN
(\S3.1.2.1) could represent the outer part of the rear jet.
   Although barely supported by the existing data, the jet hypothesis
   offers a way to explain the offset large-scale X-ray (and TeV) emission
south of the pulsar,  which cannot be associated with the shocked
pulsar wind under the assumption of wind isotropy. We speculate
that this emission
 could be produced by particles supplied through
the front jet (similar to the Vela PWN; Pavlov et al.\ 2003; Kargaltsev
\& Pavlov 2004).
 However, to account for the
luminous TeV emission ($L_{\gamma} \sim 2\times10^{-2}d_{3.5}^2
\dot{E}$), the front
 jet would have to carry
 a substantial fraction of the pulsar's spindown power,
unlike the  northwest
 jet of the Vela PWN
whose energy injection rate
 is only $\sim 10^{-3}\dot{E}$.

We cannot also exclude a possibility that the spin axis of the J1809
pulsar, and hence the PWN jets,
 are substantially misaligned with the direction of pulsar motion.
For instance, one could assume that the jets are nearly perpendicular
to the velocity vector (i.e. this vector lies in the equatorial plane; a
possible example is PSR B1706--44 and its PWN; Romani  et al.\ 2005).
In this case, the jets would be bent backwards (or even destroyed)
 by the ram pressure of the oncoming ambient medium,
while the TS in the equatorial outflow (ring-like without ram pressure
applied) would turn into an ellipse-like structure elongated in the
direction of pulsar motion, with the pulsar displaced along the major
axis toward the TS head. The observational appearance of the
shocked pulsar wind would then depend on the inclination of the
orbital plane to the line of sight, but at most inclinations we would see
an elongated structure with the pulsar shifted from the center of the
PWN in the direction of the velocity vector, in a qualitative agreement
with the observed shape of the inner PWN. Of course, such an
interpretation would imply a higher pulsar velocity than that
estimated above for a nearly isotropic pulsar wind (see eq.\ [2]). On
the other hand, in such a geometry the wind could propagate to
larger distances ahead of the pulsar
 because the CD could be easier
destroyed by various instabilities, and the pulsar wind would be
mixed with the shocked ambient medium. We note, however, that
this picture can hardly explain the large-scale X-ray emission
south-southwest of the pulsar, which, in this case, could be attributed
to the host SNR or the crushed relic PWN (see \S3.3 and \S3.4.2,
respectively).

Overall, we can conclude that
 the observed PWN morphology is generally
consistent with the assumption of the supersonic pulsar motion,
which implies a low ambient pressure (e.g., $p_{\rm amb}\ll 2\times
10^{-10} f_\Omega d_{3.5}^{-2} \sin^2i$ ergs cm$^{-3}$ for a nearly
isotropic pulsar wind). On the other hand, modeling of magnetized
anisotropic winds from fast-moving pulsars, deeper X-ray
observations, and proper motion measurements
 are needed to firmly establish the nature of the J1809 PWN
and infer its properties quantitatively.

\subsection{The pulsar}

 The spectrum of the J1809 pulsar is less certain
than that of the  PWN.
 Although the one-component PL model formally fits the spectrum,
 the fit yields a rather large $\Gamma\simeq2.6-3.2$ and suggests a
 smaller $n_{\rm H}$ than the one obtained from the PL fit to a better quality
PWN spectrum.
  An alternative description of the  pulsar spectrum is
  provided by the two-component PL+BB model that is often used to fit the spectra of young and middle-aged
   pulsars (see, e.g., KPG07a
  and references therein).
The PL component of the PL+BB fit gives $\Gamma_{\rm psr} \simeq
1.2\pm 0.6$, similar to $\Gamma_{\rm pwn}$
 and to the spectral slopes of
  other pulsars of similar ages
   (see e.g., KPG07a).
In any case, the slope is much softer than $\Gamma_{\rm psr} =
2.1 - 2.9 \dot{E}_{36}^{-1/2}\approx 0.1$ predicted by the Gotthelf's (2003)
correlation.
The X-ray efficiency of the pulsar, $\eta_{\rm psr} \equiv L_{\rm psr}/\dot{E}
\sim 2\times 10^{-5} d_{3.5}^2$ in the 0.5--8 keV band, is also not unusual
for Vela-like pulsars, as well as the ratio
$L_{\rm pwn}/L_{\rm psr} \sim 10$ (see KPG07a).

The possible thermal component of the PL+BB fit
 is
  poorly constrained, not only because of the scarce
statistics but also because the soft thermal radiation
 is strongly absorbed by the ISM. The BB temperature,
$T \sim 1.7$--2.3 MK, emitting area  $\mathcal{A} \sim 10^6$--$10^7$
m$^2$,
 and bolometric luminosity, $L_{\rm psr}^{\rm bol}
\sim (0.6$-$1.6) \times 10^{32}$ ergs s$^{-1}$, are similar to those
found from the PL+BB fits for Vela-like pulsars.  The temperature is a
factor of two higher  than the  NS surface temperatures
 predicted by standard NS cooling
models for the 50 kyr age (e.g., Yakovlev \& Pethick 2004), and the
corresponding emitting area is smaller than the NS surface area.
However, the actual spectrum of the NS thermal radiation can differ
substantially from the BB model. In particular, fitting the spectra with
hydrogen atmosphere models (Pavlov et al.\ 1995) yields lower
effective temperatures and larger emitting areas (see Pavlov et al.\
2001a for the specific example of the Vela pulsar). Unfortunately, the
quality of the data does not warrant fits with more complicated
atmosphere models.
 Unlike the temperature and the area,
 the bolometric luminosity is not so sensitive to the
  presence and
properties of the NS atmosphere. The comparison with the NS cooling
models shows that the J1809's bolometric luminosity is a factor of
$\sim 10$ lower than predicted by the ``basic'' theoretical cooling
curve ($M_{\rm NS} = 1.3 M_\odot$, no superfluidity); it is consistent
with the cooling curves for heavier NSs (e.g., $M_{\rm
NS}=1.5$--$1.6\, M_\odot$) for various superfluidity models
(Yakovlev \& Pethick 2004).

\subsection{Host SNR}

With the velocity given by Equation (2), the pulsar would have
traveled a distance of $\sim 7 n^{-1/2} d_{3.5}^{-1} f_\Omega^{1/2}
\sin i$ pc during the time equal to its spindown age, $\tau=51$ kyr.
This corresponds to a displacement of $\sim 7' n^{-1/2} d_{3.5}^{-2}
f_\Omega^{1/2} \sin^2 i$ in the plane of the sky and, for $n^{-1/2}
d_{3.5}^{-2} f_\Omega^{1/2} \sin^2 i\sim 0.6$, places the sky
projection of the pulsar's birthplace close to the apparent center of
the radio SNR G11.18+0.11 (see Fig.\ 3{\em b}), suggesting that J1809
and G11.18+0.11 were created by the same supernova explosion.
 This conjecture implies that the pulsar has overtaken the
SNR shell and left the high-pressure SNR interiors, which is consistent
with the assumption $p_{\rm amb}\ll p_{\rm ram}$, used to obtain
the pulsar velocity estimate. This could explain why J1809 looks so
different from those Vela-like PWNe that do not show cometary
morphology because they are moving subsonically in high-pressure
SNR interiors (see KPG07a for examples). However, if G11.18+0.11 is
at $d\approx 3.5$ kpc (the presumed distance to J1809), then its size,
$D\approx 8 d_{3.5}$ pc, would be surprisingly small for a 50 kyr old
SNR, unless it is expanding in an unusually dense environment. To get
more consistent SNR age and size, we have to assume a larger
distance and a smaller pulsar's true age, which would require a lower
ambient density around J1809 to match the pulsar's birthplace with
the center of G11.18+0.11, resulting in a higher inferred pulsar
velocity and Mach number. For instance, if we assume $d=5$ kpc (i.e.,
$D\sim 11$ pc) and the true age = 20 kyr, then $n\sim 0.1 f_\Omega
\sin^4i$ cm$^{-3}$ is required, which corresponds to $v\sim 300(\sin
i)^{-1}$ km s$^{-1}$, ${\mathcal M} \sim 25 T_4^{-1/2}\mu^{1/2}(\sin
i)^{-1}$. Thus, given the uncertainty of the pulsar's true age and
distance, we cannot rule out the possibility that G11.18+0.01 is the
host SNR for J1809.

It is also possible that G11.18+0.01 is just a background SNR
accidentally projected near the pulsar. In this case, the real host SNR
might be associated with the extended emission around J1809 seen in
the {\sl Chandra} and {\sl ASCA} images (Fig.\ 1, {\em bottom right}
and Fig.\ 3{\em c}). The faintness of
 the large-scale X-ray emission does not allow one to determine
its origin (thermal or nonthermal) and, therefore,
 thermal plasma emission from the SNR interior cannot be
excluded until better quality data are obtained (see \S2.2.1).
 However, the size of this putative SNR is too
small, and, more importantly, the observed PWN morphology does not
look consistent with this hypothesis. Therefore, it seems more plausible
that the extended emission is synchrotron radiation related to the PWN
(see \S3.1.2.2) rather than thermal emission from hot gas in SNR interior.
Measuring the X-ray spectrum of the extended emission in a deeper observation
would distinguish between these possibilities.

One could also speculate that the host SNR is not seen because its size
is larger than the field-of-view of our observation ($\gtrsim 20$ pc at
$d=3.5$ kpc), and  the SNR interior has become cold enough
($T\lesssim 10^5$ K) to provide ${\mathcal M}\gg 1$ and be
undetectable in X-rays. This option remains quite viable if the pulsar's
true age is close to (or exceeds) its spindown age.

\subsection{
Origin of HESS\,J1809}

TeV radiation can be produced by the inverse Compton scattering
(ICS) of low-frequency radiation (e.g., the cosmic microwave
background radiation [CMBR]) off relativistic electrons. Alternatively,
it can be generated by the $\pi^{0}\rightarrow\gamma + \gamma$
decay, the $\pi^{0}$ mesons being produced
 when relativistic protons
interact with the
 ambient matter.
Therefore, to understand the nature of a TeV source, one should
identify the {\em source} of relativistic particles and the
{\em target} with which
these particles interact.

\subsubsection{Possible targets where the TeV radiation is produced}

The omnipresent target for relativistic electrons producing
 TeV photons by the ICS
is the CMBR, with the energy density $U_{\rm CMBR} = 0.26\,\, {\rm
eV\,\, cm}^{-3}$. To produce TeV photons with energy $E_\gamma$
 by upscattering the CMBR photons
with energy $\epsilon \sim 3kT\sim 4\times 10^{-4}$ eV, electrons
with the Lorentz factor $\gamma \sim 5\times 10^7 E_{\rm
TeV}^{1/2}$ [i.e. $E_e \sim 25 E_{\rm TeV}^{1/2}$ TeV] are required,
where $E_{\rm TeV}=E_\gamma/(1\, {\rm TeV})$. As $\gamma
\epsilon \sim 20$--$100\, {\rm keV} \ll m_ec^2$ for $E_\gamma\sim
1$--20 TeV, the ICS occurs in the Thompson regime. The same
relativistic electrons produce synchrotron photons with energy
\be E_{\rm syn} \sim \gamma^2h\nu_{\rm cyc} \sim 5 \gamma_7^2
B_{-5}\,\, {\rm eV} \sim 0.1 \frac{ E_{\rm TeV}
B_{-5}}{\epsilon/(4\times 10^{-4}\, {\rm eV})}\, {\rm keV}\,, \ee
where $\gamma_7=\gamma/10^7$ and $B_{-5}=B/(10\,\mu{\rm G})$.

Close to the Galactic plane, where HESS\,J1809 is situated, a factor of
a few higher radiation energy density can be provided by IR emission
from interstellar dust and Galactic starlight.
 For instance,
 the models of interstellar radiation
field by Strong et al.\ (2000) give the energy densities of 0.6 and 2.7
eV cm$^{-3}$ for these two components, respectively, at the
galactocentric distance of $\sim 4$ kpc.  Since the ICS proceeds in the
Klein-Nishina (K.-N.) regime for $\epsilon \gtrsim 1 E_{\rm TeV}^{-1}$
eV (i.e., the K.-N. effects become important at $\epsilon \gtrsim {\rm
a\,\, few}\, \times 10^{-2}$ eV for the high-energy end of the TeV
photon spectrum), we should use a more general formula for
estimating the electron Lorentz factor:
\be \gamma \sim
10^6\left[E_{\rm TeV} + \left(E_{\rm TeV}^2 +E_{\rm
TeV}/\epsilon_{\rm eV}\right)^{1/2}\right]\,, \ee
 where
$\epsilon_{\rm eV}=\epsilon/(1\, {\rm eV})$. The electrons with such
a Lorentz factor generate synchrotron photons with energy $E_{\rm
syn} \sim 0.05 \left[E_{\rm TeV} + \left(E_{\rm TeV}^2 +E_{\rm
TeV}/\epsilon_{\rm eV}\right)^{1/2}\right]^2 B_{-5}$ eV. For instance,
electrons with $\gamma\sim 2\times 10^7$ (i.e. $E_e \sim 40$ TeV)
produce photons with $E_\gamma= 10$ TeV by upscattering
background starlight photons with a typical energy $\epsilon\sim 1$
eV, and the same electrons produce synchrotron radiation in a far-UV
range, $E_{\rm syn}\sim 6$ eV, in an interstellar field of 3 $\mu$G.
Although the radiation energy density of the starlight can substantially
exceed that of the CMBR, the emissivity is reduced by the smaller
Compton cross section in the K.-N. regime and smaller photon
number density.

The radiation energy density and photon number density (hence TeV
emissivity) can be enhanced in the vicinity of very bright sources. The
radiation energy density at an angular distance $\theta$ from a
source of radiation can be estimated as $U = F/(c\theta^2)$,
 where $F$ is the source energy flux observed at Earth;
it  exceed the ambient radiation energy
density, $U_{\rm amb}$, at
\be
\theta > (F/c U_{\rm amb})^{1/2}.
\ee

The multiwavelength data show two very bright radio-IR objects near
the center of HESS\,J1809 -- Source A and Source B (see Fig.\ 3), which
are likely molecular-dust complexes in star-forming regions. An
approximate integration of the spectral flux of Source A gives its total
flux $F_A \sim 10^{-7}$ ergs cm$^{-2}$ s$^{-1}$ (i.e. the luminosity
$L_A \sim 1.5\times 10^{38} d_{3.5}^2$ ergs s$^{-1}$). This means
that the radiation energy density near this source exceeds $U_{\rm
amb}$ at $\theta < 5' (U_{\rm amb}/1\,{\rm eV\, cm}^{-3})^{-1/2}$.
For instance, if the TeV emission is produced by the ICS of IR photons
with energies around $\epsilon \sim 0.02$ eV (which corresponds to
the maximum spectral flux of Source A), and $U_{\rm amb}\sim 0.1$
eV cm$^{-3}$ in this energy range, then $U>U_{\rm amb}$ at $\theta
\lesssim 15'$. This value is close to the observed radius of
HESS\,J1809, which suggests that the sphere of $\sim 10$ pc of the enhanced
 radiation field around Source A might
be the target where the TeV emission is produced by the ICS off the
relativistic electrons supplied by a nearby source
 [Lorentz factors $\gamma \sim
(0.5$--$6)\times 10^7$ are required to upscatter the IR photons to
$E_\gamma = 0.5$--20 TeV]. Similar estimates hold for Source B,
whose flux is a factor of 2.5 lower, and the size $\theta$ is a factor of
1.6 smaller, than those for Source A. The hypothesis that the ``photon
sphere'' of Source A or Source B
 is the site where the
TeV emission
 is produced
 via ICS
could be verified by detection of the accompanying synchrotron
radiation. However,  most  of the energy range of the
 corresponding synchrotron emission, $E_{\rm syn} \sim (1$--$200) B_{-5}$ eV,
is subject to the strong interstellar absorption.

In principle, the TeV emission could be initiated by high-energy
nucleons supplied by some source (e.g., a pulsar), which produce
decaying $\pi^0$ mesons in collisions with nucleons of the ambient
medium (e.g., Horns et al.\ 2006). The target in this case would the
nucleonic component of the circumpulsar medium, and we should
expect enhanced TeV emission in dense, cold clouds, if there are such
clouds close to the source of high-energy nucleons. The lack of an
extended IR or radio counterpart of a size similar to that of
HESS\,J1809
 (which would indicate the
presence of a large molecular cloud) suggests that the TeV source is
not associated with a localized target for high-energy nucleons.

\subsubsection{Possible sources of relativistic particles}

In the case of HESS\,J1809, the natural source of relativistic electrons
(and possibly protons) is the J1809 pulsar. In addition to that, there
are the Ch1 and Ch2 X-ray sources and the radio SNR G11.03-0.05, all
projected close the center of HESS\,J809 (see Fig.\ 3). Finally,
relativistic particles might be produced by acceleration mechanisms in
the forward shock of the putative host SNR of the J1809 pulsar.

The soft X-ray spectrum of Ch2 and
the positional coincidence with
a field star suggest that
the X-ray emission comes from an active stellar corona.
Therefore, we conclude that Ch2 is
not related to HESS\,J1809.

If the shell-like radio SNR G11.03--0.05 were a powerful source of
relativistic electrons, its synchrotron radiation would have been seen
in the X-ray range. Since no X-rays at the SNR location are detected in
our {\sl Chandra} observation, this SNR is not a viable candidate for
the source of relativistic particles that powers HESS\,J1809.

As we have discussed in \S3.1.2 and \S3.3, the faint,
 large-scale X-ray emission south of the J1809 pulsar is likely
not the host SNR of this pulsar, but it is rather related to the J1809
PWN. Moreover, TeV $\gamma$-ray emission from an SNR shock is
usually associated with the SNR shell, not the interior.
 Current radio,  X-ray and $\gamma$-ray data provide no evidence of such
a shell around J1809. Therefore,
 even if the large-scale X-ray emission belongs to the host SNR,
HESS\,J1809 is unlikely to be powered by
particles accelerated in the forward shock of this SNR.

Among the X-ray sources detected with {\sl Chandra}, Ch1 is the
closest to the projected HESS\,J1809 center (Fig.\ 3). The lack of an
IR-optical counterpart and the large lower limit on the X-ray-to-optical
flux ratio, $F_X/F_{\rm opt}\gtrsim 3$, mean that the X-ray emission
from Ch1 comes not from a usual field star. On the other hand, the
absorbing hydrogen column density, $n_{\rm H,22} = 1.2\pm 0.4$, is
close to that of the J1809 PWN, which suggests that Ch1 is a Galactic
object, not an AGN observed through the Galactic plane. The slope of
the Ch1 spectrum, $\Gamma = 1.4\pm 0.4$, suggests that it may be
either an X-ray binary or a remote PWN unresolved because of the far
off-axis location. With the optical extinction estimated from the
measured $n_{\rm H}$ value (e.g., $A_V\sim 4$--9), the limits on the
optical-NIR magnitudes (see \S2.5) virtually exclude a high mass X-ray
binary. Although a low mass X-ray binary (LMXB) in a quiescent state
could have an X-ray luminosity comparable
 to that of Ch1,
$L_{\rm Ch1}\sim 3\times 10^{32}(d/4~{\rm kpc})^{2}$ ergs s$^{-1}$,
LMXBs are not known to be sources of TeV emission.
 If Ch1 is a young pulsar with a PWN, then it might
provide relativistic particles needed for generating the TeV emission.
(We should note, however, that the TeV-to-X-ray flux ratio, $\sim
150$, would be much higher than the values, $\sim 0.01$--3, inferred
for most of the other TeV PWNe, except for the B1800--21, for which
the ratio is about 100; see Table 2 in Kargaltsev et al.\ 2007b,
hereafter KPG07b, and references therein). To assess the
likelihood
of Ch1 being the source of relativistic particles for HESS\,J1809, a
deeper {\sl Chandra} observation is needed, with Ch1 imaged close to
the telescope's optical axis.

Another plausible physical counterpart to HESS\,J1809 is the J1809
pulsar/PWN. To date, young  pulsars have been found in the vicinity
of $\sim10$ extended TeV sources (e.g., de Jager 2006; Gallant 2007),
and the likelihood of this happening by chance is very low (see, e.g.,
KPG07b). The ratio of the TeV flux of HESS~J1809 to the X-ray flux the
J1809 PWN (including the outer component of the compact PWN) is
about 100, higher than those observed in most TeV plerions (KPG07b).
However,  if one adds in the X-ray flux of the large-scale extended
emission (\S2.2.1),  then the ratio decreases down to $\sim20$ (or
even lower if one includes the diffuse emission of a larger extent,
seen by {\sl ASCA}; see Fig.\ 3{\em c}).

The center of
the brightest part of HESS~J1809 is offset from
 J1809
 by $\approx
8'$, almost in the direction of the presumed pulsar motion. Such an
offset is comparable to the $10'$--$20'$ offsets found in most of TeV
plerions (including the most secure associations
PSR\,B0833--45/HESS\,J0835--455 [Vela] and
PSR\,B1823--13/HESS\,J1825--137). An obvious explanation for such
an offset is that the photons for the ICS are provided by a compact
source (e.g., Source A) whose center lies at some distance from the
pulsar ($\geq 8 d_{3.5}$ pc for Source A). Also, the offset might be
due to the anisotropic supply of relativistic particles through the
putative front jet (see \S3.1.2).

If the TeV emission is produced by the ICS of an approximately
uniform background radiation, such as the CMBR or Galactic starlight,
then the origin of the offsets and the  asymmetries of the extended
X-ray and TeV PWN components could be attributed to the reverse
SNR shock that had propagated through the nonhomogeneous SNR
interior and reached one side of the PWN sooner than the other side,
crushing the PWN and sweeping its contents along (Blondin et al.\
2001).  One can assume that the TeV source HESS\,J1809 is powered
by this relic crushed PWN, whose relativistic electrons have not lost all
their energy to synchrotron radiation because of, e.g., a lower
magnetic field, while the compact PWN in the pulsar vicinity is created
by fresh electrons, recently injected from the pulsar magnetosphere.
As the electrons responsible for the TeV emission have been
accumulating during a substantial fraction of the pulsar's lifetime, and
the pulsar's spindown power was higher in the past, this hypothesis
could explain the uncomfortably large ratio, $\sim 0.01 d_{3.5}^2$, of
the TeV luminosity to the {\em current} spindown power. (A similar
explanation has been suggested by Aharonian et al.\ 2006 for the
PSR\,B1823--13/HESS J1825--137 association.)  This hypothesis can be
tested with deep radio observations, which should be able to detect
the synchrotron emission from the relic electrons of the crushed
PWN.

 In the above picture,  the asymmetric extended X-ray emission
south of the J1809 pulsar could also be associated with the crushed
PWN and considered as an X-ray (synchrotron) counterpart of the TeV
(IC) source HESS\,J1809. The small size of the extended X-ray
emission, in comparison with the size of the TeV source, and its offset
from the center of HESS\,J1809 could be attributed to the lower
sensitivity of the ACIS chip S2 (where the central part of HESS\,J1809 is
imaged -- see Fig.\ 3{\em a}) or to a lower magnetic field in the central
part. A deeper X-ray observation could distinguish between these two
possibilities.

The large-scale morphology of the X-ray emission around J1809 is
remarkably  similar to that around PSR B1823--13, which powers the
X-ray PWN G18.0--0.7 ($L_{X}\sim 3 \times 10^{33}$ ergs s$^{-1}$,
 angular size  $\gtrsim5'$; Gaensler et al.\ 2003). In addition to
 the extended low-surface-brightness component,
the B1823--13 PWN  has a much more compact
 ($5''$--$10''$) brighter core,
resolved by {\sl Chandra}
 (Teter et al.\ 2004).
The TeV emission from HESS~1825--137 covers an area much larger
than the X-ray PWN area, extending up to $1^{\circ}$ southward from
the pulsar (Aharonian et al.\ 2006). However, just as in the case of
PSR~J1809/HESS~J1809, both the TeV and the low-surface-brightness
X-ray emission have similar shapes, and they are offset in the same
direction with respect to the pulsar position.

Finally, we would like to point out that  there are two other young
pulsars, J1811--1925 and J1809--1943, at distances of $\approx 24'$
and $\approx 19'$, respectively, from the HESS\,J1809 center. PSR
J1811--1925 can not be related to HESS\,J1809 since it has been
associated with G11.2--0.3 (Kaspi et al.\ 2001), whose diameter,
$\approx 4'$, is
 substantially smaller than the offset from HESS~J1809.
PSR J1809--1943  is the radio counterpart of the transient Anomalous
X-ray Pulsar (AXP) XTE J1810--197 (Halpern et al.\ 2005, and
references therein).
 Its  association with HESS\,J1809
does not look very plausible because of the large spatial offset  and
because its properties
 are quite different from those of the young pulsars found in the vicinity
of other TeV sources (no TeV emission
 from AXPs have been reported so far).

Thus,
based on the current data, only two objects,
 the J1809 pulsar/PWN and, less likely, Ch1,
are plausible candidates for the source of relativistic particles
powering HESS\,J1809. In both cases, the TeV emission is generated
by ICS of either IR photons from Source A (or Source B) or CMBR
photons. To discriminate between different possibilities, the nature of
Ch1 should be established in deep, high-resolution {\sl Chandra}
observations, and the extended X-ray PWN emission should be
studied in a deep {\sl XMM-Newton} exposure. The synchrotron
emission from the alleged relic PWN could be detected in deep radio
observations.

\section{Conclusion}

  We have detected the X-ray emission from PSR B1809--19 and its synchrotron
nebula. The X-ray efficiency and spectrum of the PWN are similar to
those of many other compact PWNe, both younger and older than
J1809. The cometary shape of the compact PWN suggests that the
pulsar is moving supersonically, but no extended tail is
 detected behind the pulsar, perhaps because the Mach number is
not large enough or the exposure time is too short. Our analysis
of the PWN morphology suggests an anisotropic pulsar wind outflow,
possibly including jets  oriented along the direction of pulsar motion.
To establish the nature of the compact PWN unambiguously, proper
motion measurements and a deeper {\sl Chandra} observation are
required.

 The compact PWN is immersed in
an extended emission of lower surface brightness.  This extended
emission is offset with respect to the pulsar in the direction opposite
to that of the compact PWN (i.e.\ in the direction of the alleged proper
motion). If the extended emission is indeed powered by the
supersonically moving pulsar, then the radiating particles could be
supplied through the
 front jet of the compact PWN or brought by the
northern part of the reverse SNR shock
 that
overtook the pulsar moving southward.

The spectrum of the
 pulsar
 can be described by a two-component BB+PL model. For reasonable
 $n_{\rm H}$ values, the parameters of this
model
 and the corresponding component luminosities resemble those
 of Vela-like pulsars.

The J1809 pulsar and its PWN are located in the vicinity of the
unidentified extended TeV source HESS~J1809--193, which can be
powered by ultrarelativistic electrons accelerated in the J1809 PWN.
 In addition to J1809, we found another X-ray source, Ch1,
within the central  part of HESS\,J1809, which might also be an
unresolved pulsar/PWN. Although the true nature of Ch1 remains
elusive, we cannot rule out the possibility that Ch1 is an alternative
X-ray counterpart of HESS\,J1809. Whatever of the two objects, J1809
or Ch1, supply the relativistic electrons, the TeV emission is likely
produced by the ICS of either CMBR or IR photons from bright IRAS
sources (Source A or Source B) projected near HESS\,J1809.

\acknowledgements Our thanks are due to Divas Sanwal, PI of the {\sl
Chandra} observation, who participated in the initial analysis of the
{\sl Chandra} data. We also thank Eric Gotthelf for providing us with
the  {\sl ASCA} images with the latest calibration corrections applied.
We are grateful to Kostya Getman and Leisa Townsley for useful
discussions
 about the optical and IR properties of starforming regions.
Support for this work was provided by the National Aeronautics and
Space Administration through Chandra Award Number GO3--4075X
 issued by the Chandra X-ray Observatory Center,
which is operated by the Smithsonian Astrophysical Observatory for and
on behalf of the National Aeronautics Space Administration under
contract NAS8-03060.
This work was also partially supported by NASA grant NAG5-10865.

\end{document}